\documentstyle[]{livrev}
\RequirePackage{pslatex}

\RequirePackage{epsf}
\RequirePackage{longtable}

\begin{document}

\title{Rotating Stars in Relativity}

\author{Nikolaos Stergioulas\\
  \\
  Department of Physics, Aristotle University of Thessaloniki\\
  Thessaloniki, 54124, Greece\\
  email: niksterg@astro.auth.gr\\
  http://www.astro.auth.gr/$\sim$niksterg\\
  \\
\small{(last modified: 7 February 2003)}
}
\date{} 
\maketitle

\begin{abstract}
  Rotating relativistic stars have been studied extensively in recent
  years, both theoretically and observationally, because of the
  information one could obtain about the equation of state of matter
  at extremely high densities and because they are considered to be
  promising sources of gravitational waves. The latest theoretical
  understanding of rotating stars in relativity is reviewed in this
  updated article.  The sections on the equilibrium properties and on
  the nonaxisymmetric instabilities in $f$-modes and $r$-modes have
  been updated and several new sections have been added on analytic
  solutions for the exterior spacetime, rotating stars in LMXBs,
  rotating strange stars, and on rotating stars in numerical
  relativity.
\end{abstract}


\newpage

\section{Introduction}

Rotating relativistic stars are of fundamental interest in physics.
Their bulk properties constrain the proposed equations of state for
densities larger than nuclear density. Accreted matter in their
gravitational field undergoes high-frequency oscillations that could
become a sensitive probe for general relativistic effects.  Temporal
changes in the rotational period of millisecond pulsars can also
reveal a wealth of information about important physical processes
inside the stars or of cosmological relevance.  In addition, rotational
instabilities can produce gravitational waves, the detection of which
would initiate a new field of observational asteroseismology of
relativistic stars.

There exist several independent numerical codes for obtaining accurate
models of rotating neutron stars in full general relativity, including
one that is freely available. One recent code achieves near machine
accuracy even for uniform density models near the mass-shedding limit.
The uncertainty in the high-density equation of state still allows
numerically constructed maximum mass models to differ by as much as a
factor of two in mass, radius and angular velocity, and a factor of
eight in the moment of inertia. Given these uncertainties, an absolute
upper limit on the rotation of relativistic stars can be obtained by
imposing causality as the only requirement on the equation of state.
It then follows that gravitationally bound stars cannot rotate faster
than 0.28 ms.

In rotating stars, nonaxisymmetric perturbations have been studied in
the Newtonian and post-Newtonian approximations, in the slow-rotation
limit and in the Cowling approximation but fully relativistic
quasi-normal modes (except for neutral modes) have yet to be obtained.
A new method for obtaining such frequencies is the time-evolution of
the full set of nonlinear equations and frequencies of quasi-radial
modes have already been obtained. Time-evolutions of the linearized 
equations have also improved our understanding of the spectrum of
axial and hybrid modes in relativistic stars.

Nonaxisymmetric instabilities in rotating stars can be driven by the
emission of gravitational waves (CFS-instability) or by viscosity.
Relativity strengthens the former, while it weakens the latter.
Nascent neutron stars can be subject to the $l=2$ bar mode
CFS-instability, which would turn them into a strong gravitational
wave source. 

Axial fluid modes in rotating stars ($r$-modes) received considerable
attention since it was discovered that they are generically unstable
to the emission of gravitational waves.  The $r$-mode instability
could slow down newly-born relativistic stars and limit their spin
during accretion-induced spin-up, which would explain the absence of
millisecond pulsars with rotational periods less than $\sim1.5$ ms.
Gravitational waves from the $r$-mode instability 
could become detectable if the amplitude of $r$-modes is of order
unity.  Recent 3D simulations show that this is possible on dynamical
timescales, but nonlinear effects seem to set a much smaller
saturation amplitude on longer timescales. Still, if the signal persists
for a long time (as has been found to be the case for strange stars) 
even a small amplitude could become detectable.

Recent advances in numerical relativity have enabled the long-term
dynamical evolution of rotating stars and several interesting phenomena,
such as dynamical instabilities, pulsations modes, neutron star
and black hole formation in rotating collapse have now been studied
in full general relativity. The current studies are limited to 
relativistic polytropes, but new 3D simulations with realistic
equations of state should be expected in the near future.

The present article aims at presenting a summary of theoretical and
numerical methods that are used to describe the equilibrium properties
of rotating relativistic stars, their oscillations and their dynamical
evolution. It focuses on the most recently available preprints, in
order to rapidly communicate new methods and results.  At the end of
some sections, the reader is pointed to papers that could not be
presented in detail here or to other review articles. As new
developments in the field occur, updated versions of this article will
appear.

\section{The Equilibrium Structure of Rotating Relativistic Stars}

\subsection{Assumptions}
\label{assumptions}

A relativistic star can have a complicated structure (solid crust,
magnetic field, possible superfluid interior, possible quark core
etc.).  Still, its bulk properties can be computed with reasonable
accuracy by making several simplifying assumptions.

The matter can be modeled to be a perfect fluid because observations
of pulsar glitches have shown that the departures from a perfect fluid
equilibrium (due to the presence of a solid crust) are of order
$10^{-5}$ (see \cite{FI92}).  The temperature of a cold neutron star
does not affect its bulk properties and can be assumed to be 0~K,
because its thermal energy ($<<1$MeV $\sim 10^{10}$~K) is much smaller
than Fermi energies of the interior ($> 60$ MeV).  One can then use
a zero-temperature, {\it barotropic} equation of state (EOS) to describe
the matter:
\begin{equation}
     \epsilon = \epsilon(P),
\end{equation}
where $\epsilon$ is the energy density and $P$ is the pressure.  At birth, a
neutron star is expected to be rotating differentially, but as the
neutron star cools, several mechanisms can act to enforce uniform
rotation. Kinematical shear viscosity is acting against differential
rotation on a timescale that has been estimated to be
\cite{FI76,FI79,Cutler87}
\begin{equation}
\tau \sim 18 \times \left( \frac{\rho({\rm g/cm^3})}{10^{15}}\right)^{-5/4} 
          \left(\frac{T({\rm K})}{10^9}\right)^2
          \left(\frac{R({\rm cm})}{10^6}\right) \ \ \ {\rm yr,}
\end{equation}
where $\rho$, $T$ and $R$ are the central density, temperature and
radius of the star.  It has also been suggested that convective and
turbulent motions may enforce uniform rotation on a timescale of the
order of days \cite{Hegyi77}.  In recent work, Shapiro
\cite{Shapiro01} suggests that magnetic braking of differential
rotation by Alfv{\'e}n waves could be the most effective damping
mechanism, acting on short timescales of the order of minutes.
 
Within roughly a year after its formation, the temperature of a
neutron star becomes less than $10^9$~K and its outer core is expected
to become superfluid (see \cite{Me98} and references therein).
Rotation causes superfluid neutrons to form an array of quantized
vortices, with an intervortex spacing of
\begin{equation}
  d_n \sim 3.4 \times 10^{-3} \Omega_2^{-1/2} {\rm cm},
\end{equation}
where $\Omega_2$ is the angular velocity of the star in   $10^2~{\rm
s}^{-1}$.  On scales much 
larger than the intervortex spacing, e.g.
on the order of 1 cm, the fluid motions can be averaged and the
rotation can be considered to be uniform \cite{So87}. With such an
assumption, the error in computing the metric is of order
\begin{equation}
     \left( \frac{1 {\rm cm}}{R} \right)^2 \sim 10^{-12}, 
\end{equation}
assuming $R\sim 10$~km to be a typical neutron star radius. 

The above arguments show that the bulk properties of an isolated
rotating relativistic star can be modeled accurately by a uniformly
rotating, zero-temperature perfect fluid. Effects of differential
rotation and of finite temperature need only be considered during the
first year (or less) after the formation of a relativistic star.

\subsection{Geometry of Spacetime}

In general relativity, the spacetime geometry of a rotating star in
equilibrium can be described by a stationary and axisymmetric metric 
$g_{ab}$ of the form
\begin{equation}
  ds^2 = -e^{2 \nu} dt^2 + e^{2 \psi} (d \phi - \omega dt)^2 + e^{2 \mu }
            (dr^2+r^2 d \theta^2),    \label{e:metric}
\end{equation}
where $\nu$, $\psi$, $\omega$ and $\mu $ are four metric functions which
depend on the coordinates $r$ and $\theta$ only (see e.g. Bardeen \&
Wagoner \cite{Bardeen71}). Unless otherwise noted, we will assume
$c=G=1$. In the exterior vacuum, it is possible to reduce the number
of metric functions to three, but as long as one is interested in
describing the whole spacetime (including the source-region of
nonzero pressure), four different metric functions are required.
It is convenient to write  $e^\psi$ in the  the form
\begin{equation}
e^\psi=r \sin \theta B e^{-\nu},
\end{equation}
where $B$ is again a function of $r$ and $\theta$ only \cite{Bardeen73}.

One arrives at the above form of the metric assuming that i) the
spacetime has a timelike Killing vector field $t^a$ and a second
Killing vector field $\phi^a$ corresponding to axial symmetry, ii) the
spacetime is asymptotically flat, i.e.  $t_at^a=-1$, $\phi_a\phi^a=+\infty $
and $t_a\phi^a=0$ at spatial infinity. According to a theorem by Carter
\cite{Carter69}, the two Killing vectors commute and one can choose
coordinates $x^0=t$ and $x^3=\phi $ (where $x^a$, $a=0..3$ are the
coordinates of the spacetime), such that $t^a$ and $\phi^a$ are
coordinate vector fields. If, further, the source of the gravitational
field satisfies the circularity condition (absence of meridional
convective currents), then another theorem \cite{Carter70} shows that
the 2-surfaces orthogonal to $t^a$ and $\phi^a$ can be described by the
remaining two coordinates $x^1$ and $x^2$.  A common choice for $x^1$
and $x^2$ are {\it quasi-isotropic coordinates}, for which
$g_{r\theta}=0$, $g_{\theta \theta }=r^2 g_{rr}$ (in spherical polar coordinates),
or $g_{\varpi z }=0$, $g_{zz }=r^2 g_{\varpi \varpi }$ (in cylindrical
coordinates).  In the slow-rotation formalism by Hartle \cite{H67}, a
different form of the metric is used, requiring $g_{\theta \theta}=g_{\phi \phi }/
\sin^2 \theta$.

The three metric functions $\nu$, $\psi$ and $\omega$ can be written as
invariant combinations of the two Killing vectors $t^a$ and $\phi^a$, through
the relations
\begin{eqnarray}
t_at^a &=& g_{tt}, \\
\phi_a\phi^a &=& g_{\phi \phi}, \\
t_a\phi^a &=& g_{t\phi},
\end{eqnarray}
while the fourth metric function $\mu$ determines the conformal factor
$e^{2\mu}$ that characterizes the geometry of the orthogonal 2-surfaces.

There are two main effects that distinguish a rotating relativistic
star from its nonrotating counterpart: the shape of the star is
flattened by centrifugal forces (an effect that first appears at
second order in the rotation rate) and the local inertial frames are
dragged by the rotation of the source of the gravitational field.
While the former effect is also present in the Newtonian limit, the
latter is a purely relativistic effect. The study of the dragging of
inertial frames in the spacetime of a rotating star is assisted by the
introduction of the local Zero-Angular-Momentum-Observers (ZAMO)
\cite{Bardeen70,Bardeen73}. These are observers whose worldlines are
normal to the $t={\rm const.}$ hypersurfaces and are also called {\it
  Eulerian} observers. Then, the metric function $\omega$ is the angular
velocity of the local ZAMO with respect to an observer at rest at
infinity. Also, $e^{-\nu}$ is the time dilation factor between the
proper time of the local ZAMO and coordinate time $t$ (proper time at
infinity), along a radial coordinate line. The metric function $\psi$ has
a geometrical meaning: $e^\psi $ is the {\it proper circumferential
  radius} of a circle around the axis of symmetry. In the non-rotating
limit, the metric (\ref{e:metric}) reduces to the metric of a
non-rotating relativistic star in {\it isotropic coordinates} (see
\cite{Weinberg72} for the definition of these coordinates).

In rapidly rotating models, an {\it ergosphere} can appear, where
$g_{tt}>0$.  In this region, the rotational frame-dragging is strong
enough to prohibit counter-rotating time-like or null geodesics to
exist and particles can have negative energy with respect to a
stationary observer at infinity. Radiation fields (scalar,
electromagnetic or gravitational waves) can become unstable in the
ergosphere \cite{Friedman78}, but the associated growth time is
comparable to the age of the universe \cite{Comins78}.

The asymptotic behaviour of the metric functions $\nu$ and $\omega$ is 
\begin{eqnarray}
\nu &\sim& -\frac{M}{r} +\frac{Q}{r^3}P_2(\cos \theta), \label{nuatr} \\ 
\omega&\sim&\frac{2J}{r^3},
\end{eqnarray}
where $M$, $J$ and $Q$ are the gravitational mass, angular momentum
and quadrupole moment of the source of the gravitational field (see
Section \ref{RotEquil} for definitions). The asymptotic expansion of
the dragging potential $\omega$ shows that it decays rapidly far from the
star, so that its effect will be significant mainly in the vicinity of
the star.

\subsection{The Rotating Fluid}
When sources of non-isotropic stresses (such as a magnetic field or a
solid state of parts of the star), viscous stresses and heat transport
are neglected in constructing an equilibrium model of a relativistic
star, then the matter can be modeled as a perfect fluid, described by
the stress-energy tensor
\begin{equation}
     T^{ab} = (\epsilon+P)u^a u^b + P g^{ab},
\end{equation}
where $u^a$ is the fluid's 4-velocity. In terms of the two Killing vectors
$t^a$ and $\phi^a$, the 4-velocity can be written as
\begin{equation}
     u^a = \frac{e^{-\nu}}{\sqrt{1-v^2}} (t^a + \Omega \phi^a),
\end{equation}
where $v$ is the 3-velocity of the fluid with respect to a local ZAMO,
given by
\begin{equation}
     v = (\Omega-\omega)e^{\psi-\nu},
\end{equation}
and $\Omega\equiv u^\phi / u^t=d\phi / dt$ is the angular velocity of the fluid in the 
{\it coordinate frame}, which is equivalent to the angular velocity of
the fluid as seen by an  observer at rest at infinity. Stationary
configurations can be differentially rotating, while uniform rotation
($\Omega=$ const.) is a special case (see Section \ref{RotEquil}).

\subsection{Equations of Structure}

Having specified an equation of state of the form $\epsilon = \epsilon (P)$, the
structure of the star is determined by solving four components of
Einstein's gravitational field equations
\begin{equation}
     R_{ab} = 8 \pi \left (T_{ab}- \frac{1}{2} g_{ab}T \right),
\end{equation}
(where $R_{ab}$ is the Ricci tensor and $T=T_a{}^a$) and the equation of  
hydrostationary equilibrium. Setting $\zeta = \mu + \nu$,
one common choice for the gravitational field equations is \cite{BI76}
\begin{eqnarray}
\nabla \cdot (B \nabla \nu) &=&  \frac{1}{2} r^2\sin^2\theta B^3e^{-4\nu} \nabla\omega \cdot   \nabla\omega \label{fe1} \\ 
              && + 4 \pi B e^{2\zeta -2\nu} \left[\frac{(\epsilon+P)(1+v^2)}{1-v^2} +2P \right], \\
\nabla\cdot (r^2\sin^2\theta B^3e^{-4\nu}\nabla\omega) &=& -16 \pi r \sin \theta B^2e^{2\zeta -4 \nu}\frac{(\epsilon +P)v}{1-v^2}, \\
\nabla\cdot (r \sin \theta \nabla B) &=& 16 \pi r \sin \theta Be^{2\zeta -2\nu}P,
\end{eqnarray}
supplemented by a first-order differential equation for $\zeta$ (see
\cite{BI76}).  Above, $\nabla$ is the 3-dimensional derivative operator in
a flat 3-space with spherical polar coordinates $r, \theta, \phi$.

Thus, three of the four gravitational field equations are elliptic,
while the fourth equation is a first-order partial-differential
equation, relating only metric functions. The remaining non-zero
components of the gravitational field equations yield two more
elliptic equations and one first-order partial-differential equation,
which are consistent with the above set of four equations.

The equation of hydrostationary equilibrium follows from the
projection of the conservation of the stress-energy tensor normal to
the 4-velocity, $(\delta^c{}_b+u^cu_b)\nabla_aT^{ab}=0$, and is written as
\begin{equation}
P,_{i}+(\epsilon+P) \left[\nu,_{i}+\frac{1}{1-v^2}\left(-vv,_{i}+v^2\frac{\Omega,_{i}}{\Omega-\omega}
\right) \right] =0, \label{EulerEq}
\end{equation}
where a comma denotes partial differentiation and $i=1..3$. When the
equation of state is barotropic then the hydrostationary equilibrium
equation has a first integral of motion
\begin{equation}
\int_0^P\frac{dP}{\epsilon+P} -\ln(u^a\nabla_at) + \int_{\Omega_{\rm c}}^\Omega  F(\Omega) d \Omega = constant = \nu|_{\rm pole},
\label{Poincare}
\end{equation}
where $F(\Omega)=u_\phi u^t$ is some specifiable function of $\Omega$ only and
$\Omega_{\rm c}$ is the angular velocity on the symmetry axis.  In the
Newtonian limit, the assumption of a barotropic equation of state
implies that the differential rotation is necessarily constant on
cylinders and the existence of the integral of motion (\ref{Poincare})
is a direct consequence of the Poincar{\'e}-Wavre theorem
(which implies that when the rotation is constant on cylinders, the
effective gravity can be derived from a potential, see \cite{Tassoul78}).

\subsection{Rotation Law and Equilibrium Quantities}
\label{RotEquil}

A special case of rotation law is {\it uniform rotation} (uniform
angular velocity in the coordinate frame), which minimizes the total
mass-energy of a configuration for a given baryon number and total
angular momentum \cite{Boyer66,Hartle67}. In this case, the term
involving $F(\Omega)$ in (\ref{Poincare}) vanishes.

More generally, a simple choice of a differential-rotation law is
\begin{equation}
F(\Omega)= A^2(\Omega_{\rm c}-\Omega) = \frac{(\Omega-\omega)r^2\sin^2\theta~e^{2(\beta-\nu)}}{1-(\Omega -\omega)^2r^2\sin^2\theta~e^{2(\beta-\nu)}},
\end{equation}
where $A$ is a constant \cite{KEH89a,KEH89b}. When $A \to \infty$, the
above rotation law reduces to the uniform rotation case. In the
Newtonian limit and when $A \to 0$, the rotation law becomes a
so-called $j-$constant rotation law (specific angular momentum
constant in space), which satisfies the Rayleigh criterion for local
dynamical stability against axisymmetric disturbances ($j$ should not
decrease outwards, $dj/d\Omega<0$). The same criterion is also satisfied
in the relativistic case \cite{KEH89b}. It should be noted that
differentially rotating stars may also be subject to a shear
instability that tends to suppress differential rotation
\cite{Zahn93}.

The above rotation law is a simple choice that has proven to be
computationally convenient. More physically plausible choices must be
obtained through numerical simulations of the formation of
relativistic stars.

Equilibrium quantities for rotating stars, such as gravitational mass,
baryon mass, angular momentum etc. can be obtained as integrals over
the source of the gravitational field. A list of the most important
equilibrium quantities that can be computed for axisymmetric models,
along with the equations that define them, is displayed in Table
\ref{tab_equ}. There, $\rho$ is the rest-mass density, $u=\epsilon-\rho c^2$ is
the internal energy density, $\hat n^a= \nabla_at/|\nabla_bt\nabla^bt|^{1/2}$ is
the unit normal vector field to the $t={\rm const.}$ spacelike
hypersurfaces and $dV=\sqrt{|{}^3g|}~d^3x$ is the proper 3-volume
element (with ${}^3g$ being the determinant of the 3-metric).  It
should be noted that the moment of inertia cannot be computed directly
as an integral quantity over the source of the gravitational field. In
addition, there exists no unique generalization of the Newtonian
definition of the moment of inertia in general relativity and $I=J/
\Omega$ is a common choice.

\begin{table}
\begin{minipage}{75mm}
\caption{Equilibrium properties}
\begin{tabular}{*{2}{c}}
\\
\hline
circumferential radius & $R=e^\psi$ \\[0.5ex]
gravitational mass & $M=\int(T_{ab}-1/2 g_{ab})T t^a \hat n^b dV$ \\[0.5ex]
baryon mass  & $M_0 = \int \rho u_a \hat n^a dV$  \\[0.5ex]
internal energy & $U = \int u u_a \hat n^a dV $\\[0.5ex]

proper mass & $M_p = M_0 +U$  \\[0.5ex]

gravitational binding energy & $W=M-M_p-T$ \\[0.5ex]
angular momentum & $J=\int T_{ab} \phi^a \hat n^b dV$ \\[0.5ex]
moment of inertia & $I=J / \Omega$  \\[0.5ex]
kinetic energy & $T=1/2 J \Omega$\\[0.5ex]

\hline
\end{tabular}
\label{tab_equ}
\end{minipage}
\end{table}

\subsection{Equations of State}

\subsubsection{Relativistic Polytropes}

An analytic equation of state that is commonly used to model
relativistic stars is the adiabatic, relativistic polytropic EOS
of Tooper \cite{T65}
\begin{equation}
     P = K \rho^\Gamma,
\end{equation}
\begin{equation}
     \epsilon = \rho c^2 + \frac{P}{\Gamma-1},
\end{equation}
where $K$ and $\Gamma$ are the polytropic constant and polytropic
exponent, respectively. Notice that the above definition is {\it
different} from the form $P=K\epsilon^\Gamma$ (also due to Tooper
\cite{Tooper64}) that has also been used as a generalization of the
Newtonian polytropic EOS. Instead of $\Gamma$, one often uses the
polytropic index $N$, defined through
\begin{equation}
  \Gamma=1+\frac{1}{N}.
\end{equation}
For the above equation of state, the quantity $c^{(\Gamma-2)/(\Gamma-1)}
\sqrt{K^{1/(\Gamma-1)}/G}$ has units of length. In gravitational units
($c=G=1$), one can thus use $K^{N/2}$ as a fundamental length scale to
define dimensionless quantities. Equilibrium models are then
characterized by the polytropic index $N$ and their dimensionless
central energy density. Equilibrium properties can be scaled to
different dimensional values, using appropriate values for $K$.  For
$N<1.0$ ($N>1.0$) one obtains stiff (soft) models, while for $N\sim 0.5
- 1.0$, one obtains models with bulk properties that are comparable to
those of observed neutron star radii and masses.

Notice that for the above polytropic EOS, the polytropic index $\Gamma$
coincides with the adiabatic index of a relativistic isentropic fluid
\begin{equation}
  \Gamma =\Gamma_{\rm ad} \equiv  \frac{\epsilon+P}{P} \frac{dP}{d \epsilon}. \label{adiab}
\end{equation}
This is not the case for the polytropic equation of state
 $P= K \epsilon ^\Gamma$, which satisfies  (\ref{adiab}) only in the Newtonian
limit.

\subsubsection{Hadronic Equations of State}

The true equation of state that describes the interior of compact
stars is, still, largely unknown. This comes as a consequence of our
inability to verify experimentally the different theories that
describe the strong interactions between baryons and the many-body
theories of dense matter, at densities larger than about twice the
nuclear density (i.e. at densities larger than about $5\times 10^{14} {\rm
  gr}/{\rm cm}^3$).

Many different so-called realistic EOSs have been proposed to date
which all produce neutron star models that satisfy the currently
available observational constraints. The two most accurate
constraints are that the EOS must admit nonrotating neutron stars with
gravitational mass of at least $1.44 M_{\odot}$ and allow rotational
periods at least as small as $1.56$ ms, see \cite{PK94,KU95}. Recently,
the first direct determination of the gravitational redshift of spectral
lines produced in the neutron star photosphere has been obtained \cite{Cottam02}.
This determination (in the case of the low-mass X-ray binary EXO 0748-676)
yielded a redshift of $z=0.35$ at the surface of the neutron star, corresponding
to a mass to radius ratio of $M/R=0.23$ (in gravitational units), which is
compatible with most normal nuclear matter EOSs and incompatible
with some exotic matter EOS.

The theoretically proposed EOSs are qualitatively and quantitatively
very different from each other. Some are based on relativistic
many-body theories while others use nonrelativistic theories with
baryon-baryon interaction potentials. A classic collection of early
proposed EOSs was compiled by Arnett and Bowers \cite{AB}, while
recent EOSs are used in Salgado et al.  \cite{S94} and in
\cite{Datta98}. A review of many modern EOSs can be found in a recent
article by Haensel \cite{Haensel03}.  Detailed descriptions and tables
of several modern EOSs, especially EOSs with phase transitions, can be
found in Glendenning's book \cite{Glendenning97}.

High density equations of state with pion condensation have been
proposed by Migdal \cite{Mi71} and Sawyer and Scalapino \cite{SS73}.
The possibility of Kaon condensation is discussed by Brown and Bethe
\cite{BB94} (but see also Pandharipande et al. \cite{P95}).
Properties of rotating Skyrmion stars have been computed in
\cite{Ouyed01}.

The realistic EOSs are supplied in the form of an energy density vs.
pressure table and intermediate values are interpolated. This results
in some loss of accuracy because the usual interpolation methods do
not preserve thermodynamical consistency. Swesty \cite{S96} devised a
cubic Hermite interpolation scheme that does preserve thermodynamical
consistency and the scheme has been shown to indeed produce higher
accuracy neutron star models in Nozawa et al. \cite{N97}.

Usually, the interior of compact stars is modeled as a one-component
ideal fluid.  When neutron stars cool below the superfluid transition
temperature, the part of the star that becomes superfluid can be
described by a two-fluid model and new effects arise. Andersson and
Comer \cite{Andersson01} have recently used such a description in a
detailed study of slowly rotating superfluid neutron stars in general
relativity, while first rapidly rotating models are presented in \cite{Prix03}.

\subsubsection{Strange Quark Equations of State}
\label{s:strange_eos}

Strange quark stars are likely to exist, if the ground state of matter
at large atomic number is in the form of a quark fluid, which would
then be composed of about equal numbers of up, down and strange quarks
together with electrons, which give overall charge neutrality
\cite{Bodmer71,Farhi84}. The strangeness per unit baryon number is $\simeq
-1$.  The first relativistic models of stars composed of quark matter
were computed by Ipser, Kislinger and Morley \cite{Ipser75} and by 
Brecher and Caporaso \cite{Brecher76}, while the first
extensive study of strange quark star properties is due to Witten
\cite{Witten84}.

The strange quark matter equation of state can be represented by the 
following linear relation between pressure and energy density
\begin{equation}
P=a(\epsilon-\epsilon_0),
\label{MIT}
\end{equation}
where $\epsilon_0$ is the energy density at the surface of a bare strange
star (neglecting a possible thin crust of normal matter).  The MIT bag
model of strange quark matter involves three parameters, the bag
constant, ${\cal B}=\epsilon_0/4$, the mass of the strange quark, $m_s$, and
the QCD coupling constant, $\alpha_c$.  The constant $a$ in (\ref{MIT}) is
equal to $1/3$ if one neglects the mass of the strange quark, while it
takes the value of $a=0.289$ for $m_s=250$ MeV. When measured in units
of ${\cal B}_{60}={\cal B}/(60~{\rm MeV~fm^{-3}})$, the constant $B$
is restricted to be in the range
\begin{equation}
0.9821<{\cal B}_{60}<1.525,
\end{equation}
assuming $m_s=0$.  The lower limit is set by the requirement of
stability of neutrons with respect to a spontaneous fusion into
strangelets, while the upper limit is determined by the energy per
baryon of ${}^{56}$Fe at zero pressure (930.4 MeV).  For other values
of $m_s$ the above limits are modified somewhat.

A more recent attempt at describing deconfined strange quark matter is
the Dey et al. EOS \cite{Dey98}, which has asymptotic freedom built
in. It describes deconfined quarks at high densities and confinement
at zero pressure. The Dey et al. EOS can be approximated by a linear
relation of the same form as the MIT bag-model strange star EOS
(\ref{MIT}).  In such a linear approximation, typical values of the
constant $a$ are $0.45-0.46$ \cite{Gondek00}.

\begin{itemize}
\item {\bf Going further.} A review of strange quark star properties
  can be found in \cite{WSWG97}. Hybrid stars, that have a mixed-phase
  region of quark and hadronic matter, have also been proposed, see
  e.g. \cite{Glendenning97}. A study of the relaxation effect in
  dissipative relativistic fluid theories is presented in
  \cite{Relax}.
\end{itemize}

\subsection{Numerical Schemes}

All available methods for solving the system of equations describing
the equilibrium of rotating relativistic stars are numerical, as no
analytical self-consistent solution for both the interior and exterior
spacetime has been found.  The first numerical solutions were obtained
by Wilson \cite{W72} and by Bonazzola \& Schneider \cite{BS74}.  Here,
we will review the following methods: Hartle's slow rotation
formalism, the Newton-Raphson linearization scheme due to Butterworth
\& Ipser \cite{BI76}, a scheme using Green's functions by Komatsu et
al.  \cite{KEH89a,KEH89b}, a minimal surface scheme due to Neugebauer
\& Herold \cite{NH92}, and spectral-method schemes by Bonazzola et al.
\cite{BGSM93,BGM98} and by Ansorg et al.  \cite{Ansorg01}. Below we
give a description of each method and its various implementations
(codes).

\subsubsection{Hartle}

To order $O(\Omega^2)$ the structure of a star changes only by quadrupole
terms and the equilibrium equations become a set of ordinary
differential equations. Hartle's \cite{H67,HT68} method computes
rotating stars in this slow-rotation approximation and a review of
slowly rotating models has been compiled by Datta \cite{D88}. Weber et
al. \cite{WG91}, \cite{WGW91} also implement Hartle's formalism to
explore the rotational properties of four new EOSs.

Weber and Glendenning \cite{WG92} improve on Hartle's formalism in
order to obtain a more accurate estimate of the angular velocity at
the mass-shedding limit, but their models still show large
discrepancies compared to corresponding models computed without the
slow-rotation approximation \cite{S94}.  Thus, Hartle's formalism is
appropriate for typical pulsar (and most millisecond pulsar)
rotational periods, but it is not the method of choice for computing
models of rapidly rotating relativistic stars near the mass-shedding
limit.

\subsubsection{Butterworth and Ipser (BI)}

The BI scheme \cite{BI76} solves the four field equations following a
Newton-Raphson-like linearization and iteration procedure.  One starts
with a nonrotating model and increases the angular velocity in small
steps, treating a new rotating model as a linear perturbation of the
previously computed rotating model. Each linearized field equation is
discretized and the resulting linear system is solved. The four field
equations and the hydrostationary equilibrium equation are solved
separately and iteratively until convergence is achieved.

Space is truncated at a finite distance from the star and the boundary
conditions there are imposed by expanding the metric potentials in
powers of $1/r$. Angular derivatives are approximated by high-accuracy
formulae and models with density discontinuities are treated specially
at the surface. An equilibrium model is specified by fixing its rest
mass and angular velocity.

The original BI code was used to construct uniform density models and
polytropic models \cite{BI76,B76}. Friedman et al. \cite{FIP86,
  FIP89}, extend the BI code to obtain a large number of rapidly
rotating models based on a variety of realistic EOSs.  Lattimer et al.
\cite{L90} used a code which was also based on the BI scheme to
construct rotating stars using ``exotic'' and schematic EOSs,
including pion or Kaon condensation and strange quark matter.

\subsubsection{Komatsu, Eriguchi and Hachisu (KEH)}

In the KEH scheme \cite{KEH89a,KEH89b}, the same set of field
equations as in BI is used, but the three elliptic-type field
equations are converted into integral equations using appropriate
Green's functions. The boundary conditions at large distance from the
star are thus incorporated into the integral equations, but the region
of integration is truncated at a finite distance from the star.  The
fourth field equation is an ordinary first-order differential
equation. The field equations and the equation of hydrostationary
equilibrium are solved iteratively, fixing the maximum energy density
and the ratio of the polar radius to the equatorial radius, until
convergence is achieved. In \cite{KEH89a,KEH89b,EHN94} the original
KEH code is used to construct uniformly and differentially rotating
stars for both polytropic and realistic EOSs.

Cook, Shapiro and Teukolsky (CST) improve on the KEH scheme by
introducing a new radial variable which maps the semi-infinite region
$[0,\infty)$ to the closed region $[0,1]$. In this way, the region of
integration is not truncated and the model converges to a higher
accuracy. Details of the code are presented in \cite{CST92} and
polytropic and realistic models are computed in \cite{CST94a} and
\cite{CST94b}.

Stergioulas and Friedman (SF) implement their own KEH code following
the CST scheme. They improve on the accuracy of the code by a special
treatment of the second order radial derivative that appears in the
source term of the first-order differential equation for one of the
metric functions.  This derivative was introducing a numerical error
of $1 \% -2 \%$ in the bulk properties of the most rapidly rotating
stars computed in the original implementation of the KEH scheme. The
SF code is presented in \cite{SF95} and in \cite{SPHD}. It is
available as a public domain code, named {\tt rns}, and can be
downloaded from \cite{RNS}.

\subsubsection{Bonazzola et al. (BGSM)}

In the BGSM scheme \cite{BGSM93}, the field equations are derived in
the $3+1$ formulation. All four chosen equations that describe the
gravitational field are of elliptic type. This avoids the problem with
the second-order radial derivative in the source term of the ODE used
in BI and KEH.  The equations are solved using a spectral method, i.e.
all functions are expanded in terms of trigonometric functions in both
the angular and radial directions and a Fast Fourier Transform (FFT)
is used to obtain coefficients. Outside the star a redefined radial
variable is used, which maps infinity to a finite distance.

In \cite{S94,SAL94} the code is used to construct a large number of
models based on recent EOSs. The accuracy of the computed models is
estimated using two general relativistic Virial identities, valid for
general asymptotically flat spacetimes \cite{GB94,BG94} (see Section
\ref{s:virial}).

While the field equations used in the BI and KEH schemes assume a
perfect fluid, isotropic stress-energy tensor, the BGSM formulation
makes no assumption about the isotropy of $T_{ab}$. Thus, the BGSM
code can compute stars with a magnetic field, a solid crust or solid
interior and it can also be used to construct rotating boson stars.

\subsubsection{Lorene/rotstar}

The BGSM spectral method has been improved by Bonazzola et al.
\cite{BGM98} allowing for several domains of integration. One of the
domain boundaries is chosen to coincide with the surface of the star
and a regularization procedure is introduced for the divergent
derivatives at the surface (that appear in the density field when
stiff equations of state are used). This allows models to be computed
that are nearly free of Gibbs phenomena at the surface. The same
method is also suitable for constructing quasi-stationary models of
binary neutron stars. The new method has been used in
\cite{Gourgoulhon99} for computing models of rapidly rotating strange
stars and it has also been used in 3D computations of the onset of the
viscosity-driven instability to bar-mode formation \cite{Gondek02}.

\subsubsection{Ansorg et al. (AKM)}

A new multi-domain spectral method has been introduced in
\cite{Ansorg01,Ansorg03}. The method can use several domains inside the star,
one for each possible phase transition. Surface-adapted coordinates
are used and approximated by a two-dimensional Chebyshev-expansion.
Requiring transition conditions to be satisfied at the boundary of
each domain, the field and fluid equations are solved as a free
boundary value problem by a Newton-Raphson method, starting from an
initial guess. The field equations are simplified by using a
corotating reference frame. Applying this new method to the
computation of rapidly rotating homogeneous relativistic stars, Ansorg
et al. achieve near machine accuracy, except for configurations at the
mass-shedding limit (see Section \ref{s:compare})! The code has been used
in a systematic study of uniformly rotating homogeneous stars in general 
relativity \cite{Schoebel03}.

\subsubsection{The Virial Identities}
\label{s:virial}

Equilibrium configurations in Newtonian gravity satisfy the well-known
virial relation
\begin{equation}
2T+3(\Gamma -1)U+W=0.
\label{virial}
\end{equation}
This can be used to check the accuracy of computed numerical
solutions.  In general relativity, a different identity, valid for a
stationary and axisymmetric spacetime was found in \cite{Bonazzola73}.
More recently, two relativistic virial identities, valid for general
asymptotically flat spacetimes, have been derived by Bonazzola and
Gourgoulhon \cite{GB94,BG94}. The 3-dimensional virial identity (GRV3)
\cite{GB94} is the extension of the Newtonian virial identity
(\ref{virial}) to general relativity. The 2-dimensional (GRV2)
\cite{BG94} virial identity is the generalization of the identity
found in \cite{Bonazzola73} (for axisymmetric spacetimes) to general
asymptotically flat spacetimes. In \cite{BG94}, the Newtonian limit of
GRV2, in axisymmetry, is also derived. Previously, such a Newtonian
identity had only been known for spherical configurations
\cite{Chandrasekhar39}.

The two virial identities are an important tool for checking the
accuracy of numerical models and have been repeatedly used by several
authors \cite{BGSM93,S94,SAL94,N97,Ansorg01}.

\subsubsection{Direct Comparison of Numerical Codes}
\label{s:compare}

The accuracy of the above numerical codes can be estimated, if one
constructs exactly the same models with different codes and compares
them directly. The first such comparison of rapidly rotating models
constructed with the FIP and SF codes is presented by Stergioulas and
Friedman in \cite{SF95}.  Rapidly rotating models constructed with
several EOS's agree to $0.1 \% - 1.2 \%$ in the masses and radii and
to better than $2 \%$ in any other quantity that was compared (angular
velocity and momentum, central values of metric functions etc.). This
is a very satisfactory agreement, considering that the BI code was
using relatively few grid points, due to limitations of computing
power at the time of its implementation.

In \cite{SF95}, it is also shown that a large discrepancy between
certain rapidly rotating models, constructed with the FIP and KEH
codes, that was reported by Eriguchi et al. \cite{EHN94}, was only due
to the fact that a different version of a tabulated EOS was used in
\cite{EHN94} than by FIP.

Nozawa et al. \cite{N97} have completed an extensive direct comparison
of the BGSM, SF and the original KEH codes, using a large number of
models and equations of state. More than twenty different quantities
for each model are compared and the relative differences range from
$10^{-3}$ to $10^{-4}$ or better, for smooth equations of state. The
agreement is excellent for soft polytropes, which shows that all three
codes are correct and compute the desired models to an accuracy that
depends on the number of grid-points used to represent the spacetime.

If one makes the extreme assumption of uniform density, the agreement
is at the level of $10^{-2}$. In the BGSM code this is due to the fact
that the spectral expansion in terms of trigonometric functions cannot
accurately represent functions with discontinuous first-order
derivatives at the surface of the star. In the KEH and SF codes, the
three-point finite-difference formulae cannot accurately represent
derivatives across the discontinuous surface of the star.

The accuracy of the three codes is also estimated by the use of the
two Virial identities.  Overall, the BGSM and SF codes show a better
and more consistent agreement than the KEH code with BGSM or SF. This
is largely due to the fact that the KEH code does not integrate over
the whole spacetime but within a finite region around the star, which
introduces some error in the computed models.

    \begin{table}
           \begin{tabular}{lllllll}\hline
           \multicolumn{1}{c}{} & \multicolumn{1}{c}{AKM} & \multicolumn{1}{c}{Lorene/} &
           \multicolumn{1}{c}{SF}   & \multicolumn{1}{c}{SF}     & 
           \multicolumn{1}{c}{BGSM} & \multicolumn{1}{c}{KEH}     \\ 
           \multicolumn{2}{c}{}     & \multicolumn{1}{c}{rotstar}       &
           \multicolumn{1}{c}{(260x400)} &\multicolumn{1}{c}{(70x200)} &
           \multicolumn{1}{c}{} &\multicolumn{1}{c}{}\\ \hline
           $\bar{p}_c$& 1                  &      &        &       &       &      \\
           $r_p/r_e$&$0.7$                  &      &        &       & 1e-3  &      \\
           $\bar{\Omega}$&$1.41170848318$ & 9e-6 & 3e-4       & 3e-3  & 1e-2  & 1e-2 \\
           $\bar{M}$&$0.135798178809$     & 2e-4 & 2e-5       & 2e-3  & 9e-3  & 2e-2 \\
           $\bar{M}_0$&$0.186338658186$    & 2e-4 & 2e-4       & 3e-3  & 1e-2  & 2e-3 \\
           $\bar{R}_{circ}$&$0.345476187602$& 5e-5 & 3e-5       & 5e-4  & 3e-3  & 1e-3 \\
           $\bar{J}$&$0.0140585992949$    & 2e-5 & 4e-4       & 5e-4  & 2e-2  & 2e-2 \\
           $Z_p$&$1.70735395213$           & 1e-5 & 4e-5       & 1e-4  & 2e-2  & 6e-2 \\
  $Z_{eq}^f$&\hspace{-2.6mm}$-0.162534082217$& 2e-4 & 2e-3       & 2e-2  & 4e-2  & 2e-2 \\
          $Z_{eq}^b$&$11.3539142587$        & 7e-6 & 7e-5       & 1e-3  & 8e-2  & 2e-1 \\ \hline
          $|{\rm GRV3}|$ & $4e-13$          &  3e-6    & 3e-5     & 1e-3  & 4e-3  & 1e-1 \\    
          \hline\vspace*{2mm}
         \end{tabular}
         \caption{Detailed comparison of the accuracy of different 
numerical codes in computing a rapidly rotating, uniform density model. 
The absolute value of the relative error in each quantity, compared to 
the AKM code, is shown for the numerical codes Lorene/rotstar, SF (at two 
resolutions), BGSM and KEH (see text). The resolutions for the SF code are 
(angular $\times$ radial)  grid points. See \cite{N97} for the definition of 
the various equilibrium quantities.}
\label{Comparison}
         \end{table}                                                                                                                     

A new direct comparison of different codes is presented by Ansorg et
al. \cite{Ansorg01}.  Their multi-domain spectral code is compared to
the BGSM, KEH and SF codes for a particular uniform-density model of a
rapidly rotating relativistic star. An extension of the detailed
comparison in \cite{Ansorg01}, which includes results obtained by the
Lorene/rotstar code in \cite{Gondek02} and by the SF code with higher
resolution than the resolution used in \cite{N97}, is shown in Table
\ref{Comparison}. The comparison confirms that the virial identity
GRV3 is a good indicator for the accuracy of each code.  For the
particular model in Table \ref{Comparison}, the AKM code achieves
near double precision machine accuracy, while the Lorene/rotstar code
has typical relative accuracy of $2 \times 10^{-4}$ to $7\times 10^{-6}$ in
various quantities. The SF code at high resolution comes close to the
accuracy of the Lorene/rotstar code for this model.  Lower accuracies
are obtained with the SF, BGSM and KEH codes at the resolutions used
in \cite{N97}. 

The AKM code converges to machine accuracy when a large number of
about 24 expansion coefficients are used at a high computational cost.
With significantly fewer expansion coefficients (and comparable
computational cost to the SF code at high resolution) the achieved
accuracy is comparable to the accuracy of the Lorene/rotstar and SF
codes). Moreover, at the mass-shedding limit, the accuracy of the
AKM code reduces to about 5 digits (which is still highly accurate, of
course), even with 24 expansion coefficients, due to the nonanalytic
behaviour of the solution at the surface. Nevertheless, the AKM
method represents a great achievement, as it is the first method to
converge to machine accuracy when computing rapidly rotating stars in
general relativity.

\begin{itemize}
\item {\bf Going further.} A review of spectral methods in general
  relativity can be found in \cite{BFGM96}. A formulation for
  nonaxisymmetric, uniformly rotating equilibrium configurations in
  the second post-Newtonian approximation is presented in \cite{AS96}.
\end{itemize}

\subsection{Analytic Approximations to the Exterior Spacetime}

The exterior metric of a rapidly rotating neutron star differs
considerably from the Kerr metric. Only to lowest order in rotation,
the two metrics agree \cite{Hartle69}. At higher order, the multipole
moments of the gravitational field created by a rapidly rotating
compact star are different from the multipole moments of the Kerr
field. There have been many attempts in the past to find analytic
solutions to the Einstein equations in the stationary, axisymmetric
case, that could describe a rapidly rotating neutron star. An
interesting solution has been found recently by Manko et al.
\cite{Manko00,Manko00b}. For non-magnetized sources of zero net
charge, the solution reduces to a 3-parameter solution, involving the
mass, specific angular momentum and a parameter that depends on the
quadrupole moment of the source.  Although it depends explicitly only
on the quadrupole moment, it approximates the gravitational field of a rapidly
rotating star with higher non-zero multipole moments. It would be
interesting to determine whether this analytic quadrupole solution
approximates the exterior field of a rapidly rotating star more
accurately than the quadrupole, $O(\Omega^2)$, slow-rotation
approximation.

The above analytic solution (and an earlier one that was not
represented in terms of rational functions \cite{Manko94}) have been
used in studies of energy release during disk accretion onto a rapidly
rotating neutron star \cite{Sibgatullin98,Sibgatullin00}.  In
\cite{Shibata98}, a different approximation to the exterior spacetime,
in the form of a multipole expansion far from the star, has been used
to derive approximate analytic expressions for the location of the
innermost stable circular orbit (ISCO). Even though the analytic
solutions in \cite{Shibata98} converge slowly to an exact numerical
solution at the surface of the star, the analytic expressions for the
location and angular velocity at the ISCO are in good agreement with
numerical results.

\subsection{Properties of Equilibrium Models}

\subsubsection{Bulk Properties of Equilibrium Models}

     Neutron star models constructed with various realistic EOSs have
considerably different bulk properties, due to the large uncertainties
in the equation of state at high densities. Very compressible (soft)
EOSs produce models with small maximum mass, small radius, and large
rotation rate. On the other hand, less compressible (stiff) EOSs
produce models with a large maximum mass, large radius, and low
rotation rate. 

The gravitational mass, equatorial radius and rotational period of the
maximum mass model constructed with one of the softest EOSs (EOS B)
($1.63M_{\odot}$, 9.3km, 0.4ms) are a factor of two smaller than the
mass, radius and period of the corresponding model constructed by one
of the stiffest EOSs (EOS L) ($3.27M_{\odot}$, 18.3km, 0.8ms). The two
models differ by a factor of 5 in central energy density and a factor
of 8 in the moment of inertia!

Not all properties of the maximum mass models between proposed EOSs
differ considerably, at least not within groups of similar EOSs. For
example, most realistic hadronic EOSs predict a maximum mass model
with a ratio of rotational to gravitational energy $T/W$ of $0.11 \pm
0.02$, a dimensionless angular momentum $cJ/GM^2$ of $0.64 \pm 0.06$ and
an eccentricity of $0.66 \pm 0.04$, \cite{FI92}.  Hence, between the set
of realistic hadronic EOSs, some properties are directly related to
the stiffness of the EOS while other properties are rather insensitive
to stiffness. On the other hand, if one considers strange quark EOSs,
then for the maximum mass model, $T/W$ can become a factor of about
two larger than for hadronic EOSs.

\begin{figure}[h]

  \def\epsfsize#1#2{1#1} \centerline{\epsfbox{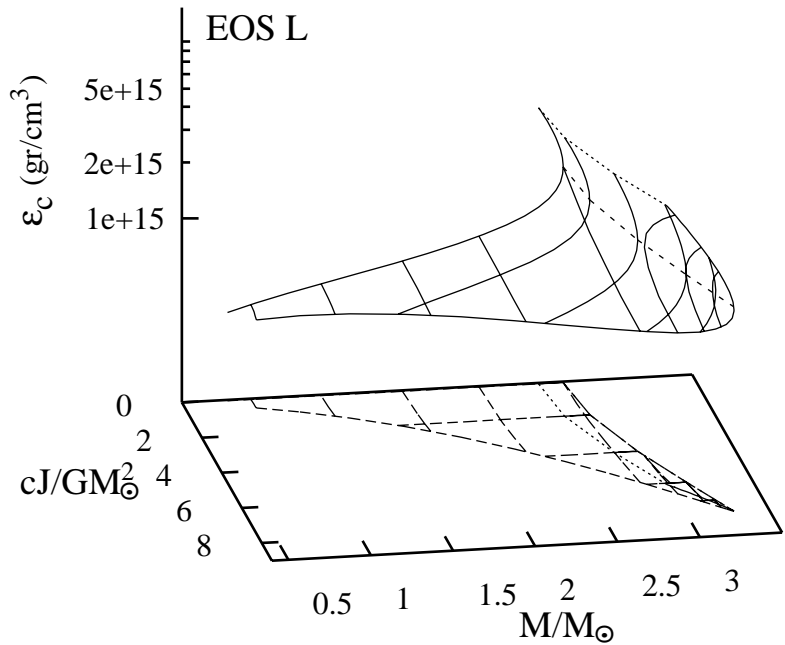}}
  \caption{\it 2-D surface of equilibrium models for EOS L. The
  surface is bounded by the nonrotating ($J=0$) and mass-shedding
  ($\Omega=\Omega_K$) limits and formed by constant $J$ and constant $M_0$
  sequences (solid lines). The projection of these sequences in the
  $J$-$M$ plane are shown as long-dashed lines. Also shown are the
  axisymmetric instability sequence (short-dashed line). The
  projection of the 2-D surface in the $J$-$M$ plane shows an
  overlapping (see dotted lines). (Figure 7 of Stergioulas and
  Friedman, ApJ \cite{SF95}.)}  \label{fig_equilsurface}
\end{figure}

Compared to nonrotating stars, the effect of rotation is to increase
the equatorial radius of the star and also to increase the mass that
can be sustained at a given central energy density. As a result, the
mass of the maximum mass rotating model is roughly $15 \% -20 \%$
higher than the mass of the maximum mass nonrotating model, for
typical realistic hadronic EOSs.  The corresponding increase in radius
is $30 \% -40 \%$. The effect of rotation in increasing the mass and
radius becomes more pronounced in the case of strange quark EOSs (see
Section \ref{s:strange}).

The deformed shape of a rapidly rotating star creates a distortion,
away from spherical symmetry, in its gravitational field. Far from the
star, the dominant multipole moment of the rotational distortion is
measured by the quadrupole-moment tensor $Q_{ab}$. For uniformly
rotating, axisymmetric and equatorially symmetric configurations, one
can define a scalar quadrupole moment $Q$, which can be extracted from
the asymptotic expansion, at large r, of the metric function $\nu$, as
in Equation (\ref{nuatr}).

Laarakkers \& Poisson \cite{LP97} numerically compute the scalar
quadrupole moment $Q$ for several equations of state, using the
rotating neutron star code {\tt rns} \cite{RNS}. They find that for
fixed gravitational mass $M$, the quadrupole moment is given as a
simple quadratic fit
\begin{equation}
      Q = -a \frac{J^2}{ Mc^2},
\end{equation}
where $J$ is the angular momentum of the star and $a$ is a
dimensionless quantity that depends on the equation of state. The
above quadratic fit reproduces $Q$ with a remarkable accuracy. The
quantity $a$ varies between $a \sim 2$ for very soft EOSs and $a \sim 8$
for very stiff EOSs, for $M=1.4 M_{\odot}$ neutron stars. This is
considerably different from a Kerr black hole, for which $a=1$
\cite{Thorne80}.

For a given zero-temperature EOS, the uniformly rotating equilibrium
models form a 2-dimensional surface in the 3-dimensional space of
central energy density, gravitational mass and angular momentum
\cite{SF95}, as shown in Figure \ref{fig_equilsurface} for EOS L.  
The surface is limited by the nonrotating models ($J=0$) and by the
models rotating at the mass-shedding (Kepler) limit, i.e.  at the
maximum allowed angular velocity so that the star does not shed mass
at the equator.  Cook et al. \cite{CST92,CST94a,CST94b} have shown
that the model with maximum angular velocity does not coincide with
the maximum mass model, but is generally very close to it in central
density and mass. Stergioulas and Friedman \cite{SF95} show that the
maximum angular velocity and maximum baryon mass equilibrium models
are also distinct. The distinction becomes significant in the case
where the EOS has a large phase transition near the central density of
the maximum mass model; otherwise the models of maximum mass, baryon
mass, angular velocity and angular momentum can be considered to
coincide for most purposes.

\begin{itemize}
\item {\bf Going further.} Although rotating relativistic stars are
  nearly perfectly axisymmetric, a small degree of asymmetry (e.g.
  frozen into the solid crust during its formation) can become a
  source of gravitational waves. A recent review of this can be found
  in \cite{Jones01}.
\end{itemize}

\subsubsection{Mass-shedding Limit and the Empirical Formula}

Mass-shedding occurs when the angular velocity of the star reaches the
angular velocity of a particle in a circular Keplerian orbit at the
equator, i.e.
\begin{equation}
  \Omega = \Omega_K,
\end{equation} 
where
\begin{equation}
  \Omega_K=\frac{\omega'}{2\psi'} + e^{\nu-\psi}\left[c^2\frac{\nu'}{\psi'}+\left(\frac{\omega'}{2\psi'}e^{\psi-\nu}\right)^2
                               \right]^{1/2}+\omega.
\end{equation}
In differentially rotating stars, even a small amount of differential
rotation can increase the angular velocity required for mass-shedding
significantly. Thus, a newly-born, hot, differentially rotating
neutron star or a massive compact object created in a binary neutron
star merger, could be sustained (temporarily) in equilibrium by
differential rotation, even if a uniformly rotating configuration with
same rest mass does not exist.

In the Newtonian limit the maximum angular velocity of uniformly
rotating polytropic stars is approximately $\Omega_{max} \simeq (2/3)^{3/2}
(GM/R^3)^{1/2}$ (this is derived using the Roche model, see
\cite{ST83}).  For relativistic stars, the empirical formula
\cite{HZ89,FIP89,Fr89}
\begin{equation}
\Omega_{max} = 0.67 \sqrt{\frac{G M_{max}}{R_{max}^3}},
                 \label{e:empirical}
\end{equation}
gives the maximum angular velocity in terms of the mass and radius of
the maximum mass {\em nonrotating} model with an accuracy of $5 \% -7
\%$, without actually having to construct rotating models. A revised
empirical formula, using a large set of EOSs, has been computed in
\cite{Haensel95}.

The empirical formula results from universal proportionality relations
that exist between the mass and radius of the maximum mass rotating
model and those of the maximum mass nonrotating model for the same
EOS. Lasota et al. \cite{LHA96} find that, for most EOSs, the
coefficient in the empirical formula is an almost linear function of
the parameter
\begin{equation}
\chi_s = \frac{2GM_{max}}{R_{max} c^2}.
\end{equation}
The Lasota et al. empirical formula
\begin{equation}
\Omega_{max} = {\cal C} (\chi_s)  \sqrt{\frac{G M_{max}}{R_{max}^3}},
                 \label{e:lasota}
\end{equation}
with ${\cal C} (\chi_s)=0.468+0.378 \chi_s$, reproduces the exact values with
a relative error of only $1.5 \%$.

Weber and Glendenning \cite{WG91,WG92}, derive analytically a similar
empirical formula in the slow rotation approximation. However, the
formula they obtain involves the mass and radius of the maximum mass
{\it rotating} configuration, which is different from what is involved in
(\ref{e:empirical}).

\subsubsection{Upper Limits on Mass and Rotation: Theory vs. Observation}

The maximum mass and minimum period of rotating relativistic stars
computed with realistic hadronic EOSs from the Arnett and Bowers
collection \cite{AB} are about $3.3 M_{\odot}$ (EOS L) and $0.4$ ms (EOS
B), while $1.4 M_{\odot}$ neutron stars, rotating at the Kepler limit,
have rotational periods between $0.53$ ms (EOS B) and $1.7$ ms (EOS M)
\cite{CST94b}. The maximum, accurately measured, neutron star mass is
currently still $1.44 M_{\odot}$ (see e.g. \cite{Kerkwijk95}, but there
are also indications for $2.0 M_{\odot}$ neutron stars \cite{KFC97}.
Core-collapse simulations have yielded a bi-modal mass distribution of
the remnant, with peaks at about $1.3M_\odot$ and $1.7M_\odot$
\cite{Timmes96} (the second peak depends on the assumption for the
high-density EOS - if a soft EOS is assumed, then black hole formation
of this mass is implied). Compact stars of much higher mass, created
in a neutron star binary merger, could be temporarily supported
against collapse by strong differential rotation \cite{Baumgarte00}.

When magnetic-field effects are ignored, conservation of angular
momentum can yield very rapidly rotating neutron stars at birth.
Recent simulations of the rotational core collapse of evolved rotating
progenitors \cite{Heger00,Fryer00} have demonstrated that rotational
core collapse can easily result in the creation of neutron stars with
rotational periods of the order of 1 ms (and similar initial rotation
periods have been estimated for neutron stars created in the
accretion-induced collapse of a white dwarf \cite{Liu01}). The
existence of a magnetic field may complicate this picture. Spruit \&
Phinney \cite{Spruit98} have presented a model in which a strong
internal magnetic field couples the angular velocity between core and
surface during most evolutionary phases. The core rotation decouples
from the rotation of the surface only after central carbon depletion
takes place. Neutron stars born in this way would have very small
initial rotation rates, even smaller than the ones that have been
observed in pulsars associated with supernova remnants. In this model,
an additional mechanism is required to spin-up the neutron star to
observed periods. On the other hand, Livio \& Pringle \cite{Livio98}
argue for a much weaker rotational coupling between core and surface
by a magnetic field, allowing for the production of more rapidly
rotating neutron stars than in \cite{Spruit98}.  A new investigation
by Heger et al., yielding intermediate initial rotation rates, is
presented in \cite{Heger03}.  Clearly, more detailed computations are
needed to resolve this important question.

The minimum observed pulsar period is still 1.56ms \cite{KU95}, which
is close to the experimental sensitivity of most pulsar searches. New
pulsar surveys, in principle sensitive down to a few tenths of a
millisecond, have not been able to detect a sub-millisecond pulsar
\cite{DA96,Damico00,Crawford00,Edwards01}.  This not too surprising,
as there are several explanations for the absence of sub-millisecond
pulsars. In one model, the minimum rotational period of pulsars could
be set by the occurrence of the $r$-mode instability in accreting
neutron stars in LMXB's \cite{Andersson00}. Other models are based on
the standard magnetospheric model for accretion-induced spin-up
\cite{White97} or on the idea that gravitational radiation (produced
by accretion-induced quadrupole deformations of the deep crust)
balances the spin-up torque \cite{Bildsten98,Ushomirsky00}. It has
also been suggested \cite{Burderi01} that the absence of
sub-millisecond pulsars in all surveys conducted so far is because
they could be more likely to be found only in close systems (of
orbital period $P_{orb}\sim 1$ hr), for which the current pulsar surveys
are still lacking the required sensitivity. The absence of
sub-millisecond pulsars in wide systems is suggested to be due to the
turning-on of the accreting neutron stars as pulsars, in which case
the pulsar wind is shown to halt further spin-up.

\begin{itemize}
  \item {\bf Going further.} A review by J. L. Friedman on the
  upper limit on rotation of relativistic stars can be found in 
  \cite{Fr95}.
\end{itemize}

\subsubsection{The Upper Limit on Mass and Rotation Set by Causality}

If one is interested in obtaining upper limits on the mass and
rotation rate, independently of the proposed EOSs, one has to rely on
fundamental physical principles.  Instead of using realistic EOSs, one
constructs a set of schematic EOSs that satisfy only a minimal set of
physical constraints, which represent what we know about the equation
of state of matter with high confidence. One then searches among all
these EOSs to obtain the one that gives the maximum mass or minimum
period. The minimal set of constraints that have been used in such
searches are that
\begin{enumerate}
   \item the high density EOS matches to the known low density EOS at
         some matching energy density $\epsilon_m$,
   \item the matter at high densities satisfies the causality
         constraint (the speed of sound is less than the speed of light).
\end{enumerate}
In relativistic perfect fluids, the speed of sound is the
characteristic velocity of the evolution equations for the fluid and
the causality constraint translates into the requirement
\begin{equation}
           dp/d \epsilon \leq 1.
\end{equation}
(see \cite{GL}). It is assumed that the 
fluid will still behave as a perfect fluid when it is perturbed 
from equilibrium.

For nonrotating stars, Rhoades and Ruffini showed that the EOS that
satisfies the above two constraints and yields the maximum mass
consists of a high density region as stiff as possible (i.e. at the
causal limit, $dp/d \epsilon=1$), that matches directly to the known low
density EOS. For a chosen matching density $\epsilon_m$, they computed a
maximum mass of $3.2 M_{\odot}$. However, this is not the theoretically
maximum mass of nonrotating neutron stars, as is often quoted in the
literature. Hartle and Sabbadini \cite{HS77} point out that $M_{max}$
is sensitive to the matching energy density and Hartle \cite{H78}
computes $M_{max}$ as a function of $\epsilon_m$.
\begin{equation}
     M_{max} = 4.8 \ \Bigl( \frac{2 \times 10^{14} {\rm gr/cm}^3} 
                   { \epsilon_m} \Bigr)^{1/2} M_{\odot}.
\end{equation}

In the case of rotating stars, Friedman and Ipser \cite{FI87} assume
that the absolute maximum mass is obtained by the same EOS as in the
nonrotating case and compute $M_{max}$ as a function of matching
density, assuming the BPS EOS holds at low densities. A more recent
computation \cite{KSF97} uses the FPS EOS at low densities, arriving
at a similar result as in \cite{FI87}
\begin{equation}
     M^{rot}_{max} = 6.1 \ \Bigl( \frac{2 \times 10^{14} {\rm gr/cm}^3} 
                   { \epsilon_m} \Bigr)^{1/2} M_{\odot},
\end{equation}
where, $2 \times 10^{14} {\rm gr/cm}^3$ is roughly nuclear saturation 
density for the FPS EOS.

A first estimate of the absolute minimum period of uniformly rotating,
gravitationally bound stars was computed by Glendenning \cite{G92} by
constructing nonrotating models and using the empirical formula
(\ref{e:empirical}) to estimate the minimum period.  Koranda,
Stergioulas and Friedman \cite{KSF97} improve on Glendenning's results
by constructing accurate rapidly rotating models and show that
Glendenning's results are accurate to within the accuracy of the
empirical formula.

Furthermore, they show that the EOS satisfying the minimal set of
constraints and yielding the minimum period star consists of a high
density region at the causal limit (CL EOS), $P=(\epsilon -\epsilon_C)$, (where
$\epsilon_C$ is the lowest energy density of this region), which is matched
to the known low density EOS through an intermediate constant pressure
region (that would correspond to a first-order phase transition).
Thus, the EOS yielding absolute minimum period models is as stiff as
possible at the central density of the star (to sustain a large enough
mass) and as soft as possible in the crust, in order to have the
smallest possible radius (and rotational period).

The absolute minimum period of uniformly rotating stars is an (almost
linear) function of the maximum observed mass of nonrotating neutron
stars
\begin{equation}
    P_{min}= 0.28 {\rm ms} + 0.2 (M_{max}^{nonrot.}/M_\odot - 1.44) {\rm ms},
\label{e:pmin}
\end{equation}
and is rather insensitive to the matching density $\epsilon_m$ (the above
result was computed for a matching number density of $0.1 {\rm
  fm}^{-3}$).  In \cite{KSF97}, it is also shown that an absolute
limit on the minimum period exists even without requiring that the EOS
matches to a known low density EOS, i.e. if the CL EOS $P=(\epsilon -\epsilon_C)$
terminates at a surface energy density of $\epsilon_C$. This is not so for
the causal limit on the maximum mass. Thus, without matching to a
low-density EOS, the causality limit on $P_{min}$ is lowered by only
$3 \%$, which shows that the currently known part of the nuclear EOS
plays a negligible role in determining the absolute upper limit on the
rotation of uniformly rotating, gravitationally bound stars.

The above results have been confirmed in \cite{Haensel99}, where it is
shown that the CL EOS has $\chi_s=0.7081$, independent of $\epsilon_C$, and
the empirical formula (\ref{e:lasota}) reproduces the numerical result
(\ref{e:pmin}) to within 2\%.

\subsubsection{Supramassive Stars and Spin-Up Prior to Collapse}

Since rotation increases the mass that a neutron star of given central
density can support, there exist sequences of neutron stars with
constant baryon mass that have no nonrotating member. Such sequences
are called supramassive, as opposed to normal sequences that do have a
nonrotating member. A nonrotating star can become supramassive by
accreting matter and spinning-up to large rotation rates; in another
scenario, neutron stars could be born supramassive after a core
collapse. A supramassive star evolves along a sequence of constant
baryon mass, slowly losing angular momentum. Eventually, the star
reaches a point where it becomes unstable to axisymmetric
perturbations and collapses to a black hole.

In a neutron star binary merger, prompt collapse to a black hole 
can be avoided if the equation of state is sufficiently stiff and/or
the equilibrium is supported by strong differential rotation. The
maximum mass of differentially rotating supramassive neutron stars
can be significantly larger than in the case of uniform rotation. A
detailed study of this mass-increase has recently appeared in \cite{Lyford03}.

Cook et al.  \cite{CST92,CST94a,CST94b} have discovered that a
supramassive relativistic star approaching the axisymmetric
instability, will actually spin up before collapse, even though it
loses angular momentum.  This, potentially observable, effect is
independent of the equation of state and it is more pronounced for
rapidly rotating massive stars. Similarly, stars can spin up by loss
of angular momentum near the mass-shedding limit, if the equation of
state is extremely stiff or extremely soft. 

If the equation of state features a phase transition to e.g. quark
matter, then the spin-up region is very large and most millisecond
pulsars (if supramassive) would need to be spinning up \cite{Spyrou02}
- the absence of spin-up in known millisecond pulsars indicates that
either large phase transitions do not occur, or that the equation of
state is sufficiently stiff, so that millisecond pulsars are not
supramassive.

\subsubsection{Rotating Magnetized Neutron Stars}

The presence of a magnetic field has been ignored in the models of
rapidly rotating relativistic stars that were considered in the
previous sections. The reason is that the observed surface dipole
magnetic field strength of pulsars ranges between $10^8$ G and $2 \times
10^{13}$ G. These values of the magnetic field strength imply a
magnetic field energy density that is too small compared to the energy
density of the fluid, to significantly affect the structure of a
neutron star. However, one cannot exclude the existence of neutron
stars with higher magnetic field strengths or the possibility that
neutron stars are born with much stronger magnetic fields, which then
decay to the observed values (of course, there are also many arguments
against magnetic field decay in neutron stars \cite{PK94}).  In
addition, even though moderate magnetic field strengths do not alter
the bulk properties of neutron stars, they may have an effect on the
damping or growth rate of various perturbations of an equilibrium
star, affecting its stability.  For these reasons, a fully
relativistic description of magnetized neutron stars is desirable and,
in fact, Bocquet et al. \cite{BBGN95} achieved the first numerical
computation of such configurations. Here we give a brief summary of
their work:

A magnetized relativistic star in equilibrium can be described by the
coupled Einstein-Maxwell field equations for stationary, axisymmetric
rotating objects with internal electric currents. The stress-energy
tensor includes the electromagnetic energy density and is
non-isotropic (in contrast to the isotropic perfect fluid
stress-energy tensor). The equilibrium of the matter is given not only
by the balance between the gravitational force, centrifugal force and
the pressure gradient, but the Lorentz force due to the electric
currents also enters the balance. For simplicity, Bocquet et al.
consider only poloidal magnetic fields, which preserve the circularity
of the spacetime.  Also, they only consider stationary configurations,
which excludes magnetic dipole moments non-aligned with the rotation
axis, since in that case the star emits electromagnetic and
gravitational waves.  The assumption of stationarity implies that the
fluid is necessarily rigidly rotating (if the matter has infinite
conductivity) \cite{BGSM93}.  Under these assumptions, the
electromagnetic field tensor $F^{ab}$ is derived from a potential
four-vector $A_a$ with only two non-vanishing components, $A_t$ and
$A_{\phi}$, which are solutions of a scalar Poisson and a vector Poisson
equation respectively. Thus, the two equations describing the
electromagnetic field are of similar type as the four field equations
that describe the gravitational field.

For magnetic field strengths larger than about $10^{14}$ G, one
observes significant effects, such as a flattening of the equilibrium
configuration.  There exists a maximum value of the magnetic field
strength, of the order of $10^{18}$ G, for which the magnetic field
pressure at the center of the star equals the fluid pressure. Above
this value no stationary configuration can exist.

A strong magnetic field allows a maximum mass configuration with
larger $M_{max}$ than for the same EOS with no magnetic field and this
is in analogy with the increase of $M_{max}$ induced by rotation. For
nonrotating stars, the increase in $M_{max}$, due to a strong magnetic
field, is $13 \% - 29 \%$, depending on the EOS.  Correspondingly, the
maximum allowed angular velocity, for a given EOS, also increases in
the presence of a strong magnetic field.

Another application of general-relativistic E/M theory in neutron
stars is the study of the evolution of the magnetic field during
pulsar spin-down.  A detailed analysis of the evolution equations of
the E/M field in a slowly rotating magnetized neutron star has
revealed that effects due to the spacetime curvature and due to the
rotational frame-dragging are present in the induction equations, when
one assumes finite electrical conductivity (see \cite{Rezzolla01} and
references therein). Numerical solutions of the evolution equations of
the E/M have shown, however, that for realistic values of the
electrical conductivity, the above relativistic effects are small,
even in the case of rapid rotation \cite{Rezzolla02}.

\begin{itemize}
\item {\bf Going further.} An $O(\Omega)$ slow-rotation approach for the
  construction of rotating magnetized relativistic stars is presented in
  \cite{Gupta98}.
\end{itemize}

\subsubsection{Rapidly Rotating Proto-Neutron Stars}

Following the gravitational collapse of a massive stellar core, a
proto-neutron star (PNS) is born. Initially it has a large radius of
about 100 km and a temperature of 50-100 MeV. The PNS may be born with a
large rotational kinetic energy and initially it will be
differentially rotating. Due to the violent nature of the
gravitational collapse, the PNS pulsates heavily, emitting significant
amounts of gravitational radiation. After a few hundred pulsational
periods, bulk viscosity will damp the pulsations significantly.  Rapid
cooling due to deleptonization transforms the PNS into a hot neutron
star of $T \sim 10$ MeV, shortly after its formation. In addition,
viscosity or other mechanisms (see Sec. \ref{assumptions}) enforce
uniform rotation and the neutron star becomes quasi-stationary.  Since
the details of the PNS evolution determine the properties of the
resulting cold NSs, proto-neutron stars need to be modeled realistically
in order to understand the structure of cold neutron stars.

\begin{figure}[h]
  \def\epsfsize#1#2{0.6#1}
  \centerline{\epsfbox{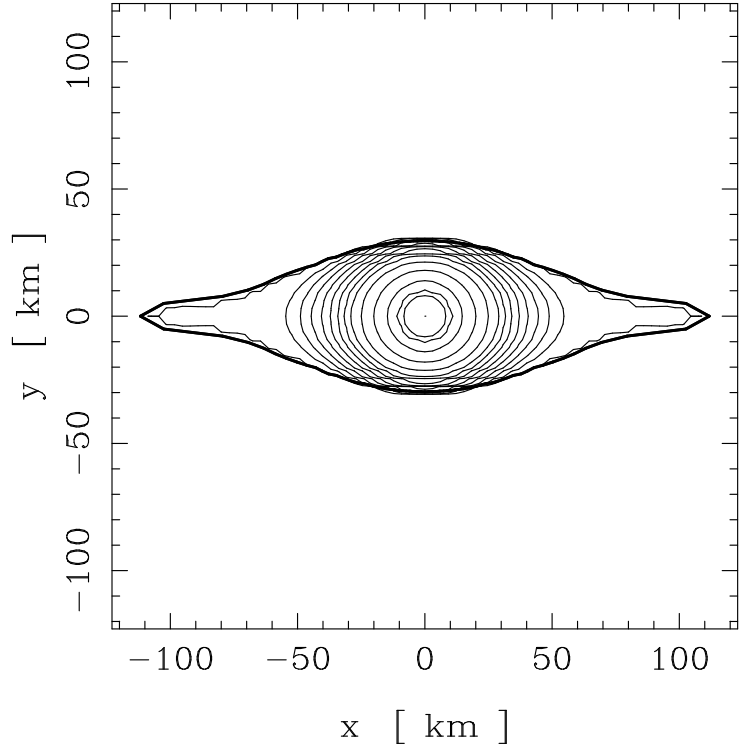}} \caption{\it
  Iso-energy density lines of a differentially rotating proto-neutron
  star at the mass-shedding limit, of rest mass $M_0=1.5~{\rm
  M}_\odot$. (Figure 5a of Goussard, Haensel and Zdunik, A\&A
  \cite{GHZ98}; used with permission.)}  \label{fig_protroNS}
\end{figure}

Hashimoto et al. \cite{HOE94} and Goussard et al. \cite{GHZ96}
construct fully relativistic models of rapidly rotating, hot
proto-neutron stars. The authors use finite-temperature EOSs
\cite{O93,LS91}, to model the interior of PNSs. Important (but largely
unknown) parameters, that determine the local state of matter, are the
lepton fraction $Y_l$ and the temperature profile. Hashimoto et al.
consider only the limiting case of zero lepton fraction $Y_l=0$ and
classical isothermality, while Goussard et al. consider several
non-zero values for $Y_l$ and two different limiting temperature
profiles - a constant entropy profile and a relativistic isothermal
profile. In both \cite{HOE94} and \cite{O93}, differential rotation is
neglected to a first approximation.

The construction of numerical models with the above assumptions shows
that, due to the high temperature and the presence of trapped
neutrinos, PNSs have a significantly larger radius than cold NSs.
These two effects give the PNS an extended envelope which, however,
contains only roughly $0.1 \%$ of the total mass of the star. This
outer layer cools more rapidly than the interior and becomes
transparent to neutrinos, while the core of the star remains hot and
neutrino opaque for a longer time. The two regions are separated by
the ``neutrino-sphere''.

Compared to the $T=0$ case, an isothermal EOS with temperature of
25MeV has a maximum mass model of only slightly larger mass. In
contrast, an isentropic EOS with a nonzero trapped lepton number
features a maximum mass model that has a considerably lower mass than
the corresponding model in the $T=0$ case and a stable PNS transforms
to a stable neutron star. If, however, one considers the hypothetical
case of a large amplitude phase transition which softens the cold EOS
(such as a Kaon condensate), then $M_{max}$ of cold neutron stars is
lower than $M_{max}$ of PNSs and a stable PNS with maximum mass will
collapse to a black hole after the initial cooling period. This
scenario of delayed collapse of nascent neutron stars has been
proposed by Brown and Bethe \cite{BB94} and investigated by Baumgarte
et al. \cite{BST96}.

An analysis of radial stability of PNSs \cite{GHZ97} shows that, for
hot PNSs, the maximum angular velocity model almost coincides with the
maximum mass model, as is also the case for cold EOSs.

Because of their increased radius, PNSs have a different mass-shedding
limit than cold NSs. For an isothermal profile, the mass-shedding
limit proves to be sensitive to the exact location of the neutrino
sphere. For the EOSs considered in \cite{HOE94} and \cite{GHZ96} PNSs
have a maximum angular velocity that is considerably less than the
maximum angular velocity allowed by the cold EOS. Stars that have
nonrotating counterparts (i.e. that belong to a normal sequence)
contract and speed up while they cool down. The final star with
maximum rotation is thus closer to the mass-shedding limit of cold
stars than was the hot PNS with maximum rotation.  Surprisingly, stars
belonging to a supramassive sequence exhibit the opposite behavior.
If one assumes that a PNS evolves without losing angular momentum or
accreting mass, then a cold neutron star produced by the cooling of a
hot PNS has a smaller angular velocity than its progenitor. This
purely relativistic effect was pointed out in \cite{HOE94} and
confirmed in \cite{GHZ96}.

It should be noted here, that a small amount of differential rotation
significantly affects the mass-shedding limit, allowing more massive
stars to exist than uniform rotation allows. Taking differential
rotation into account, Goussard et al. \cite{GHZ98} suggest that
proto-neutron stars created in a gravitational collapse cannot spin
faster than 1.7 ms. A similar result has been obtained by Strobel et
al. \cite{Strobel99}. The structure of a differentially rotating
proto-neutron star at the mass-shedding limit is shown in Figure
\ref{fig_protroNS}. The outer layers of the star form an extended
disk-like structure.

The above stringent limits on the initial period of neutron stars are
obtained assuming that the PNS evolves in a quasi-stationary manner
along a sequence of equilibrium models. It is not clear whether these
limits will remain valid, if one studies the early evolution of PNS
without the above assumption. It is conceivable that the thin hot
envelope surrounding the PNS does not affect the dynamics of the bulk
of the star. If the bulk of the star rotates faster than the
(stationary) mass-shedding limit of a PNS model, then the hot envelope
will simply be shed away from the star in the equatorial region, if it
cannot remain bounded to the star even when differentially
rotating. Such a fully dynamical study is needed to obtain an accurate
upper limit on the rotation of neutron stars.

\begin{itemize}
\item {\bf Going further.} The thermal history and evolutionary tracks
  of rotating PNS (in the 2nd-order slow-rotation approximation) have
  been studied recently in \cite{Sumiyoshi99}.

\end{itemize}

\subsubsection{Rotating Strange Quark Stars}
\label{s:strange}

\begin{figure}[p]
  \def\epsfsize#1#2{0.6#1}
  \centerline{\epsfbox{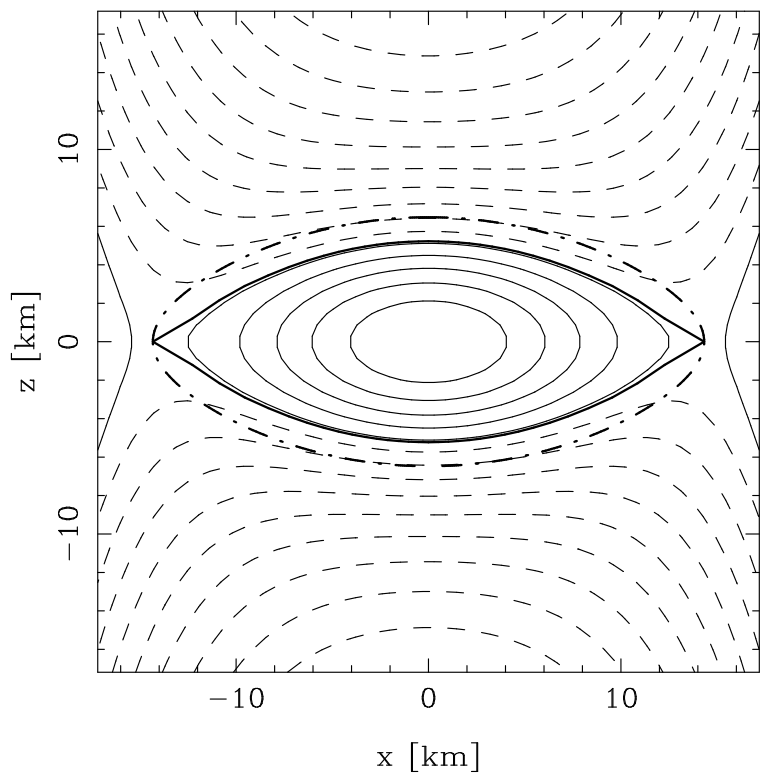}} \caption{\it Meridional
  plane cross section of a rapidly rotating strange star at the
  mass-shedding limit, obtained with a multi-domain spectral code.
  The various lines are isocontours of the log-enthalpy $H$, as
  defined in \cite{Gourgoulhon99}.  Solid lines indicate a positive
  value of $H$ and dashed lines a negative value (vacuum). The thick
  solid line denotes the stellar surface.  The thick dot-dashed line
  denotes the boundary between the two computational domains.  (Figure
  4 of Gourgoulhon, Haensel, Livine, Paluch, Bonazzola and Marck
  in A\&A \cite{Gourgoulhon99}; used with permission.)}  \label{fig:strange}
\end{figure}

Most rotational properties of strange quark stars differ considerably
from the properties of rotating stars constructed with hadronic EOSs.
First models of rapidly rotating strange quark stars were computed by
Friedman (\cite{Friedman89}, quoted in
\cite{Glendenning89a,Glendenning89b}) and by Lattimer et al.
\cite{L90}.  Colpi \& Miller \cite{Colpi92} use the $O(\Omega^2)$
approximation and find that the spin of strange stars (newly-born, or
spun-up by accretion) may be limited by the CFS-instability to the
$l=m=2$ $f$-mode, since rapidly rotating strange stars tend to have
$T/W > 0.14$. Rapidly rotating strange stars at the mass-shedding
limit have been computed first by Gourgoulhon et
al. \cite{Gourgoulhon99} and the structure of a representative
model is displayed in Figure \ref{fig:strange}.

Nonrotating strange stars obey scaling relations with the
constant ${\cal B}$ in the MIT bag-model of the strange quark matter EOS
(Section \ref{s:strange_eos}) and Gourgoulhon et al.
\cite{Gourgoulhon99} also obtain scaling relations for the model with
maximum rotation rate.  The maximum angular velocity scales as
\begin{equation}
\Omega_{\rm max}=9.92 \times 10^3 \sqrt{{\cal B}_{60}} \ {\rm s}^{-1},
\end{equation}
while the allowed range of ${\cal B}$ implies an allowed range of $0.513 \ 
{\rm ms}<P_{\rm min}<0.640 \ {\rm ms}$.  The empirical formula
(\ref{e:empirical}) also holds for rotating strange stars with an
accuracy of better than $2\% $. A derivation of the empirical formula
in the case of strange stars, starting from first principles, has been
presented by Cheng \& Harko \cite{Cheng00}, who found that some
properties of rapidly rotating strange stars can be reproduced by
approximating the exterior spacetime by the Kerr metric.

Since both the maximum mass nonrotating and maximum mass rotating
models obey similar scalings with $B$, the ratios
\begin{equation}
\frac{M_{\rm max}^{\rm rot}}{M_{\rm max}^{\rm stat}}=1.44,   \ \ \ \ \ \ \ \
\frac{R_{\rm max}^{\rm rot}}{R_{\rm max}^{\rm stat}}=1.54, 
\end{equation}
are independent of ${\cal B}$ (where $R_{\rm max}$ is the radius of the
maximum mass model).  The maximum mass increases by 44\% and the
radius of the maximum mass model by 54\%, while the corresponding
increase for hadronic stars is, at best, $\sim20$\% and $\sim40$\%,
correspondingly.  The rotational properties of strange star models
which are based on the Dey et al. EOS \cite{Dey98} are similar to
those of the MIT bag-model EOS \cite{Bodmer71,Witten84,Farhi84}, but
some quantitative differences exist \cite{Gondek00}.

Accreting strange stars in LMXBs will follow different evolutionary
paths in a mass vs.  central energy density diagram, than accreting
hadronic stars \cite{Zdunik01b}. When (and if) strange stars reach the
mass-shedding limit, the ISCO still exists \cite{Stergioulas99} (while
it disappears for most hadronic EOSs). Stergioulas, Klu{\'z}niak and
Bulik \cite{Stergioulas99} show that the radius and location of the
ISCO for the sequence of mass-shedding models also scales as
${\cal B}^{-1/2}$, while the angular velocity of particles in circular orbit
at the ISCO scales as ${\cal B}^{1/2}$.  Additional scalings with the
constant $a$ in the strange quark EOS (that were proposed in
\cite{L90}) are found to hold within an accuracy of better than $\sim
9\%$ for the mass-shedding sequence
\begin{equation}
M \propto a^{1/2}, \ \ \ \ R \propto a^{1/4}, \ \ \ \ \Omega \propto a^{-1/8}.
\end{equation}
In addition, it is found that models at mass-shedding can have $T/W$
as large as $0.28$ for $M=1.34 \ M_\odot$.

As strange quark stars are very compact, the angular velocity at the
ISCO can become very large.  If the $1066 \ {\rm Hz}$ upper QPO
frequency in 4U 1820-30 (see \cite{Kaaret99} and references therein)
is the frequency at the ISCO, then it rules out most models of slowly-rotating
strange stars in LMXBs.  However, in \cite{Stergioulas99} it is shown
that rapidly rotating bare strange stars are still compatible with
this observation, as they can have ISCO frequencies $<1 \ {\rm kHz}$
even for $1.4 \ M_\odot $ models.  On the other hand, if strange stars
have a thin solid crust, the ISCO frequency at the mass-shedding limit
increases by about 10\% (compared to a bare strange star of the same
mass) and the above observational requirement is only satisfied for
slowly rotating models near the maximum nonrotating mass, assuming
some specific values of the parameters in the strange star EOS
\cite{Zdunik00,Zdunik01}.  Moderately rotating strange stars, with
spin frequencies around 300Hz can also be accommodated for some values
of the coupling constant $\alpha_c$ \cite{Zdunik00b} (see also
\cite{Gondek01} for a detailed study of the ISCO frequency for
rotating strange stars). The $1066 \ {\rm Hz}$ requirement for the
ISCO frequency depends, of course, on the adopted model of kHz QPOs in
LMXBs and other models exist (see next section).

If strange stars can have a solid crust, then the density at the
bottom of the crust is the neutron drip density $\epsilon_{\rm ND}\simeq 4.1 \times
10^{11} \ {\rm g} \ {\rm cm}^{-3}$, as neutrons are absorbed by
strange quark matter. A strong electric field separates the nuclei of
the crust from the quark plasma. In general, the mass of the crust
that a strange star can support is very small, of the order of
$10^{-5}M_\odot$. Rapid rotation increases by a few times the mass of the
crust and the thickness at the equator becomes much larger than the
thickness at the poles \cite{Zdunik01}. Zdunik, Haensel \& Gourgoulhon
\cite{Zdunik01} also find that the mass $M_{\rm crust}$ and thickness
$t_{\rm crust}$ of the crust can be expanded in powers of the spin
frequency $\nu_3=\nu/(10^3 \ {\rm Hz})$ as
\begin{eqnarray}
M_{\rm crust} &=& M_{\rm crust,0}(1+0.24 \nu_3^2+0.16 \nu_3^8), \\
t_{\rm crust} &=& t_{\rm crust,0} (1+0.4 \nu_3^2+0.3\nu_3^6),
\end{eqnarray}
where a subscript ``0'' denotes nonrotating values. For $\nu\leq 500$ Hz,
the above expansion agrees well with the quadratic expansion derived
previously by Glendenning \& Weber \cite{Glendenning92}. In a spinning
down magnetized strange quark star with crust, part of the crust will
gradually dissolve into strange quark matter, in a strongly exothermic
process. In \cite{Zdunik01}, it is estimated that the heating due to
deconfinement may exceed the neutrino luminosity from the core of a
strange star older than $\sim 1000$ years and may therefore influence
the cooling of this compact object (see also \cite{Yuan99}).

\subsection{Rotating Relativistic Stars in LMXBs}

\subsubsection{Particle orbits and kHz quasi-periodic oscillations}

In the last few years, X-ray observations of accreting sources in Low
Mass X-ray Binaries (LMXBs) have revealed a rich phenomenology that is
waiting to be interpreted correctly and could lead to significant
advances in our understanding of compact objects (see
\cite{Lamb98,VanDerKlis00,Psaltis01}). The most important feature of these
sources is the observation of (in most cases) twin kHz quasi-periodic
oscillations (QPOs) . The high frequency of these variabilities and
their quasi-periodic nature are evidence that they are produced in
high-velocity flows near the surface of the compact star. To date,
there exist a large number of different theoretical models that
attempt to explain the origin of these oscillations. No consensus has
been reached, yet, but once a credible explanation is found, it will
lead to important constraints on the properties of the compact object
that is the source of the gravitational field in which the kHz
oscillations take place. The compact stars in LMXBs are spun up by
accretion, so that many of them may be rotating rapidly and the correct inclusion
of rotational effects in the theoretical models for kHz QPOs is
important. Under simplifying assumptions for the angular momentum and
mass evolution during accretion, one can use accurate rapidly rotating
relativistic models to follow the possible evolutionary tracks of
compact stars in LMXBs \cite{Cook94,Zdunik01b}.

In most theoretical models, one or both kHz QPO frequencies are
associated with the orbital motion of inhomogeneities or blobs in a
thin accretion disk. In the actual calculations, the frequencies are
computed in the approximation of an orbiting test particle, neglecting
pressure terms. For most equations of state, stars that are massive
enough posses an innermost-stable circular orbit (ISCO) and the
orbital frequency at the ISCO has been proposed to be one of the two
observed frequencies. To first order in the rotation rate, the orbital
frequency at the prograde ISCO is given by (see Klu{\'z}niak, Michelson
and Wagoner \cite{Kluzniak90})
\begin{equation}
f_{\rm ISCO} \simeq 2210 (1+0.75j) \left(\frac{1 M_\odot}{M} \right) {\rm Hz},
\end{equation}
where $j=J/M^2$. At larger rotation rates, higher order contributions
of $j$ as well as contributions from the quadrupole moment $Q$ become
important and an approximate expression has been derived by Shibata
and Sasaki \cite{Shibata98}, which, when written as above and
truncated to the lowest order contribution of $Q$ and to $O(j^2)$,
becomes
\begin{equation}
f_{\rm ISCO} \simeq 2210 (1+0.75j+0.78j^2-0.23Q_2) \left(\frac{1 M_\odot}{M} \right) {\rm Hz},
\end{equation}
where $Q_2=-Q/M^3$.

Notice that, while rotation increases the orbital frequency at the
ISCO, the quadrupole moment has the opposite effect, which can become
important for rapidly rotating models. Numerical evaluations of
$f_{\rm ISCO}$ for rapidly rotating stars have been used in
\cite{Miller98} to arrive at constraints on the properties of the
accreting compact object.

In other models, orbits of particles that are eccentric and slightly
tilted with respect to the equatorial plane are involved.  For
eccentric orbits, the periastron advances with a frequency $\nu_{PA}$
that is the difference between the Keplerian frequency of azimuthal
motion $\nu_K$ and the radial epicyclic frequency $\nu_r$. On the other
hand, particles in slightly tilted orbits fail to return to the
initial displacement $\psi$ from the equatorial plane, after a full
revolution around the star.  This introduces a nodal precession
frequency $\nu_{PA}$, which is the difference between $\nu_K$ and the
frequency of the motion out of the orbital plane (meridional
frequency) $\nu_{\psi}$.  Explicit expressions for the above frequencies,
in the gravitational field of a rapidly rotating neutron star, have
been derived recently by Markovi{\'c} \cite{Markovic00}, while in
\cite{Markovic00b} highly eccentric orbits are considered. Morsink and
Stella \cite{Morsink99} compute the nodal precession frequency for a
wide range of neutron star masses and equations of state and (in a
post-Newtonian analysis) separate the precession caused by the
Lense-Thirring (frame-dragging) effect from the precession caused by
the quadrupole moment of the star.  The nodal and periastron
precession of inclined orbits have also been studied using an
approximate analytic solution for the exterior gravitational field of
rapidly rotating stars \cite{Sibgatullin02}. These precession
frequencies are relativistic effects and have been used in
several models to explain the kHz QPO frequencies
\cite{Stella99,Psaltis00a,Abramowicz01,Kluzniak02,Amsterdamski02}.

It is worth mentioning that it has recently been found that an ISCO
also exists in Newtonian gravity, for models of rapidly rotating
low-mass strange stars. The instability in the circular orbits is
produced by the large oblateness of the star
\cite{Kluzniak01,Zdunik01c,Amsterdamski02}.

\subsubsection{Angular Momentum Conservation During Burst Oscillations}

Some sources in LMXBs show signatures of type I X-ray bursts, which
are thermonuclear flashes on the surface of the compact star
\cite{Lewin95}. Such bursts show nearly-coherent oscillations in the
range 270-620~Hz (see \cite{VanDerKlis00,Strohmayer01} for recent
reviews). One interpretation of the burst oscillations is that they
are the result of rotational modulation of surface asymmetries during
the burst. In such case, the oscillation frequency should be nearly
equal to the spin frequency of the star. This model currently has
difficulties in explaining some observed properties, such as the
oscillations seen in the tail of the burst, the frequency increase
during the burst, the need for two anti-podal hot spots in some
sources that ignite at the same time etc. Alternative models also
exist (see e.g.  \cite{Psaltis00}).

In the spin-frequency interpretation, the increase in the oscillation
frequency by a few Hz during the burst is explained as follows: The
burning shell is supposed to first decouple from the neutron star and
then gradually settle down onto the surface. By angular momentum
conservation, the shell spins-up, giving rise to the observed
frequency increase. Cumming et al. \cite{Cumming01} compute the
expected spin-up in full general relativity and taking into account
rapid rotation. Assuming that the angular momentum per unit mass
is conserved, the change in angular velocity with radius is given by
\begin{equation}
\frac{d\ln \Omega}{d\ln r} = - 2 \left[ \left( 1-\frac{v^2}{2}-\frac{R}{2}\frac{\partial \nu}{\partial r}
\right) \left(1-\frac{\omega}{\Omega}\right) - \frac{R}{2\Omega}\frac{\partial \omega}{\partial r} \right],
\end{equation}
where $R$ is the equatorial radius of the star and all quantities are
evaluated at the equator. The slow-rotation limit of the above 
result was derived previously by Abramowicz et al. \cite{Abramowicz01}.
The fractional change in angular velocity
during spin-up can then be estimated as
\begin{equation}
\frac{\Delta \Omega}{\Omega}= \frac{d\ln \Omega}{d\ln r}\left( \frac{\Delta r}{R} \right),
\end{equation}
where $\Delta r$ is the coordinate expansion of the burning shell, a
quantity that depends on the shell's composition. Cumming et al. find
that the spin down expected if the atmosphere rotates rigidly is a
factor of two to three times smaller than observed values. More
detailed modeling is needed to fully explain the origin and properties
of burst oscillations.

\begin{itemize}
\item {\bf Going further.} A very interesting topic is the modeling of
  the expected X-ray spectrum of an accretion disk in the
  gravitational field of a rapidly rotating neutron star as it could
  lead to observational constraints on the source of the gravitational
  field, see e.g.
  \cite{Thampan98,Sibgatullin98,Sibgatullin00,Bhattacharyya02,Bhattacharyya02b},
  where work initiated by Kluzniak and Wilson \cite{Kluzniak91} in the
  slow-rotation limit is extended to rapidly rotating relativistic
  stars.
\end{itemize}

\section{Oscillations and Stability}

The study of oscillations of relativistic stars is motivated by the
prospect of detecting such oscillations in electromagnetic or
gravitational wave signals. In the same way that helioseismology is
providing us with information about the interior of the Sun, the
observational identification of oscillation frequencies of
relativistic stars could constrain the high-density equation of state
\cite{Ko97}. The oscillations could be excited after a core collapse
or during the final stages of a neutron star binary merger.  Rapidly
rotating relativistic stars can become unstable to the emission of
gravitational waves.

When the oscillations of an equilibrium star are of small magnitude
compared to its radius, it will suffice to approximate them as linear
perturbations. Such perturbations can be described in two equivalent
ways.  In the Lagrangian approach, one studies the changes in a given
fluid element as it oscillates about its equilibrium position.  In the
Eulerian approach, one studies the change in fluid variables at a
fixed point in space.  Both approaches have their strengths and
weaknesses.

In the Newtonian limit, the Lagrangian approach has been used to
develop variational principles \cite{LBO67,FS78} but the Eulerian
approach proved to be more suitable for numerical computations of mode
frequencies and eigenfunctions \cite{IM85,M85,IL90,IL91a,IL91b}.
Clement \cite{C81} used the Lagrangian approach to obtain axisymmetric
normal modes of rotating stars, while nonaxisymmetric solutions were
obtained in the Lagrangian approach by Imamura et al.  \cite{IFD85}
and in the Eulerian approach by Managan \cite{M85} and Ipser and
Lindblom \cite{IL90}. While a lot has been learned from Newtonian
studies, in the following we will focus on the relativistic treatment
of oscillations of rotating stars.

\subsection{Quasi-Normal Modes of Oscillation}

A general linear perturbation of the energy density in a static and
spherically symmetric relativistic star can be written as a sum of
quasi-normal modes that are characterized by the indices $(l,m)$ of
the spherical harmonic functions $Y_l^m$ and have angular and
time-dependence of the form
\begin{equation}
   \delta \epsilon   \sim f(r) P_l^m(\cos \theta ) e^{i(m\phi+\omega_i t)},
\end{equation}
where $\delta$ indicates the Eulerian perturbation of a quantity,
$\omega_i$ is the angular frequency of the mode, as measured by a distant
inertial observer, $f(r)$ represents the radial dependence of the
perturbation and $P_l^m(\cos \theta)$ are the associated Legendre
polynomials. Normal modes of nonrotating stars are degenerate in $m$
and it suffices to study the axisymmetric $(m=0)$ case.

The Eulerian perturbation in the fluid 4-velocity, $\delta u^a$, can be
expressed in terms of vector harmonics, while the metric perturbation,
$\delta g_{ab}$, can be expressed in terms of spherical, vector and tensor
harmonics.  These are either of "polar" or "axial" parity.  Here,
parity is defined to be the change in sign under a combination of
reflection in the equatorial plane and rotation by $\pi$. A polar
perturbation has parity $(-1)^l$, while an axial perturbation has
parity $(-1)^{l+1}$.  Because of the spherical background, the polar
and axial perturbations of a nonrotating star are completely
decoupled.

A normal mode solution satisfies the perturbed gravitational field
equations
\begin{equation}
            \delta(G^{ab}-8 \pi T^{ab})=0, \label{dGab}
\end{equation}
and the perturbation of the conservation of the stress-energy tensor
\begin{equation}
            \delta(\nabla_aT^{ab})=0,
\end{equation}
with suitable boundary conditions at the center of the star and at infinity. 
The latter equation is decomposed into an equation for the perturbation in the
energy density $\delta \epsilon$ and into equations for the three spatial
components of the perturbation in the 4-velocity, $\delta u^a$. As linear
perturbation have a gauge freedom, at most six components of the
perturbed field equations (\ref{dGab}) need to be considered.

For a given pair $(l,m)$, a solution exists for any value of the
frequency $\omega_i$, consisting of a mixture of ingoing- and
outgoing-wave parts. Outgoing quasi-normal modes are defined by the
discrete set of eigenfrequencies for which there are no incoming waves
at infinity.  These are the modes that will be excited in various
astrophysical situations.

The main modes of pulsation that are known to exist in relativistic
stars have been classified as follows ($f_0$ and $\tau_0$ are typical
frequencies and damping times of the most important modes in the
nonrotating limit):
\begin{enumerate}
  \item {\it Polar fluid modes}
    
    Are slowly damped modes analogous to the Newtonian fluid
    pulsations:
   \begin{itemize}  
   \item $f$(undamental)-mode: surface mode due to the interface
     between the star and its surroundings ($f_0 \sim 2$kHz,
     $\tau_0<1$s),

    \item  $p$(ressure)-modes: nearly radial ($f_0 >4$kHz, 
           $\tau_0 > 1$s),
           
         \item $g$(ravity modes): nearly tangential, only exist in
           stars that are non-isentropic or that have a composition
           gradient or first-order phase transition ($f_0<500$Hz, $\tau_0
           > 5$s).
    \end{itemize} \item {\it Axial and Hybrid fluid modes}

  \begin{itemize}  
    
  \item {\it inertial} modes: degenerate at zero frequency in
    nonrotating stars. In a rotating star, some inertial modes are
    generically unstable to the CFS-instability; frequencies from zero to
    kHz, growth times inversely proportional to a high power of the
    star's angular velocity. Hybrid inertial modes have both axial and
    polar parts even in the limit of no rotation.
    
  \item $r$(otation) modes: a special case of inertial modes that
    reduce to the classical axial $r$-modes in the Newtonian limit.
    Generically unstable to the CFS-instability with growth times as
    short as a few seconds at high rotation rates.

   \end{itemize}  

  \item {\it Polar and axial spacetime modes}

   \begin{itemize}  
   \item $w$(ave)-modes: Analogous to the quasi-normal modes of a
     black hole (very weak coupling to the fluid).  High frequency,
     strongly damped modes ($f_0>6$kHz, $\tau_0 \sim 0.1$ ms).
   \end{itemize}  
\end{enumerate}
For a more detailed description of various types of oscillation modes
see \cite{KB97,KLH96,MacD88,CAR86,KScha98}.

\subsection{Effect of Rotation on Quasi-Normal Modes}

In a continuous sequence of rotating stars that includes a nonrotating
member, a quasi-normal mode of index $l$ is defined as the mode which,
in the nonrotating limit, reduces to the quasi-normal mode of the same
index $l$. Rotation has several effects on the modes of a corresponding
nonrotating star:
\begin{enumerate}
\item The degeneracy in the index $m$ is removed and a nonrotating
  mode of index $l$ is split into $2l+1$ different $(l,m)$ modes.
  
\item Prograde ($m<0$) modes are now different from retrograde ($m>0$)
  modes.
  
\item A rotating ``polar" $l$-mode consists of a sum of purely polar
  and purely axial terms \cite{SPHD}, e.g. for $l=m$
\begin{equation}
         P_l^{rot} \sim \sum_{l'=0}^\infty(P_{l+2l'} +A_{l+2l' \pm 1}),
\end{equation}
that is, rotation couples a polar $l$-term to an axial $l \pm 1$ term
(the coupling to the $l+1$ term is, however, strongly favoured over
the coupling to the $l-1$ term \cite{CF91}).  Similarly, for a
rotating ``axial'' mode with $l=m$
\begin{equation}
         A_l^{rot} \sim \sum_{l'=0}^\infty(A_{l+2l'} +P_{l+2l' \pm 1}),
\end{equation}
\item Frequencies and damping times are shifted. In general,
  frequencies (in the inertial frame) of prograde modes increase,
  while those of retrograde modes decrease with increasing rate of
  rotation.
\item In rapidly rotating stars, {\it apparent intersections} between
  higher-order modes of different $l$ can occur. In such cases, the
  shape of the eigenfunction is used in the mode classification.
\end{enumerate} 

In rotating stars, quasi-normal modes of oscillation have only been
studied in the slow-rotation limit, in the post-Newtonian and in the
Cowling Approximations. The solution of the fully-relativistic
perturbation equations for a rapidly rotating star is still a very
challenging task and only recently they have been solved for zero-
frequency (neutral) modes \cite{SPHD,SF97}. First frequencies of
quasi-radial modes have now been obtained through 3D numerical
time-evolutions of the nonlinear equations \cite{Font02}.

\begin{itemize}
\item {\bf Going further.} The equations that describe oscillations of
  the solid crust of a rapidly rotating relativistic star are derived
  by Priou in \cite{Pr92}. The effects of superfluid hydrodynamics on
  the oscillations of neutron stars have been investigated by several
  authors, see e.g. \cite{LM94,Comer99,Andersson01b,Andersson02} and
  references therein.

\end{itemize}

\subsection{Axisymmetric Perturbations}

\subsubsection{Secular and Dynamical Axisymmetric Instability}

Along a sequence of nonrotating relativistic stars with increasing
central energy density, there is always a model for which the mass
becomes maximum. The maximum-mass turning point marks the onset of an
instability in the fundamental radial pulsation mode of the star.

Applying the turning point theorem provided by Sorkin \cite{So82},
Friedman, Ipser and Sorkin \cite{FIS88} show that in the case of
rotating stars a secular axisymmetric instability sets in when the
mass becomes maximum along a sequence of constant angular momentum. An
equivalent criterion (implied in \cite{FIS88}) is provided by Cook et al. \cite{CST92} : the
secular axisymmetric instability sets in when the angular momentum
becomes minimum along a sequence of constant rest mass.  The
instability first develops on a secular timescale that is set by the
time required for viscosity to redistribute the star's angular
momentum. This timescale is long compared to the dynamical timescale
and comparable to the spin-up time following a pulsar glitch.
Eventually, the star encounters the onset of dynamical instability and
collapses to a black hole (see \cite{Shibata00c} for recent numerical
simulations). Thus, the onset of the secular instability to
axisymmetric perturbations separates stable neutron stars from neutron
stars that will collapse to a black hole.

Goussard et al. \cite{GHZ96} extend the stability criterion to hot
protoneutron stars with nonzero total entropy. In this case, the loss
of stability is marked by the configuration with minimum angular
momentum along a sequence of both constant rest mass and total
entropy.  In the nonrotating limit, Gondek et al. \cite{GHZ97} compute
frequencies and eigenfunctions of radial pulsations of hot
proto-neutron stars and verify that the secular instability sets in at
the maximum mass turning point, as is the case for cold neutron stars.

\subsubsection{Axisymmetric Pulsation Modes}

\begin{figure}[p]
  \def\epsfsize#1#2{0.5#1} \centerline{\epsfbox{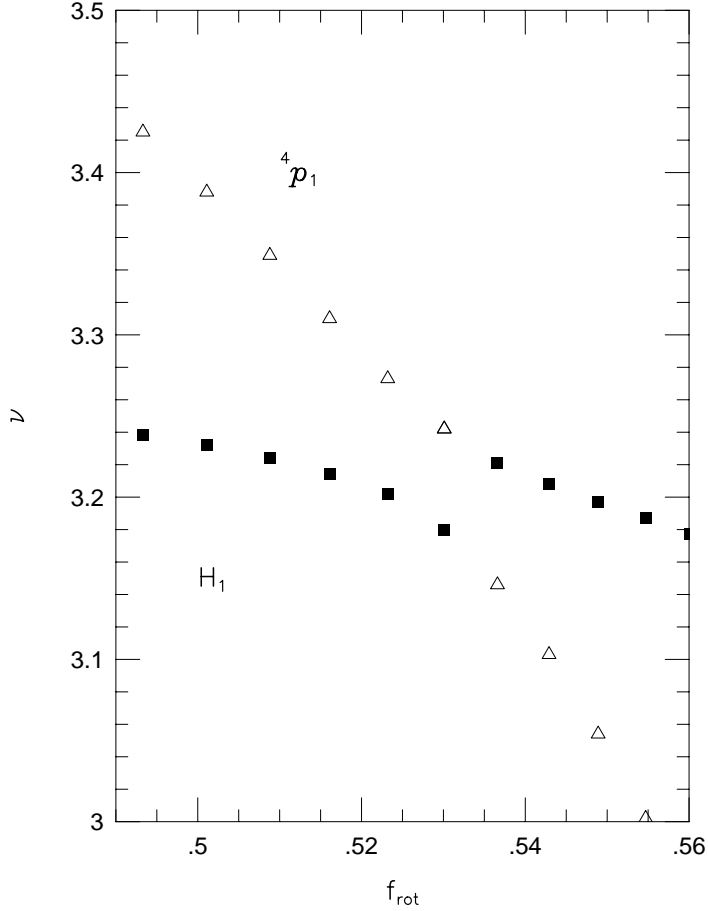}}
  \caption{\it Apparent intersection (due to avoided crossing) of the
  axisymmetric first quasi-radial overtone ($H_1$) and the first
  overtone of the $l=4$, $p$-mode (in the Cowling approximation).
  Frequencies are normalized by $\sqrt{\rho_c/4\pi}$, where $\rho_c$ is the
  central energy density of the star.  The rotational frequency
  $f_{rot} $ at the mass-shedding limit is $0.597$ (in the above
  units). Along continuous sequences of computed frequencies, mode
  eigenfunctions are exchanged at the avoided crossing. Defining
  quasi-normal mode sequences by the shape of their eigenfunction, the
  $H_1$ sequence (filled boxes) appears to intersect with the
  ${}^4p_1$ sequence (triangle), but each sequence shows a
  discontinuity, when the region of apparent intersection is
  well-resolved. (Figure 3 of Yoshida and Eriguchi, MNRAS
  \cite{Yoshida00}; used with permission.)}  \label{fig_apparent}
\end{figure}

\begin{figure}[p]
  \def\epsfsize#1#2{0.9#1}
  \centerline{\epsfbox{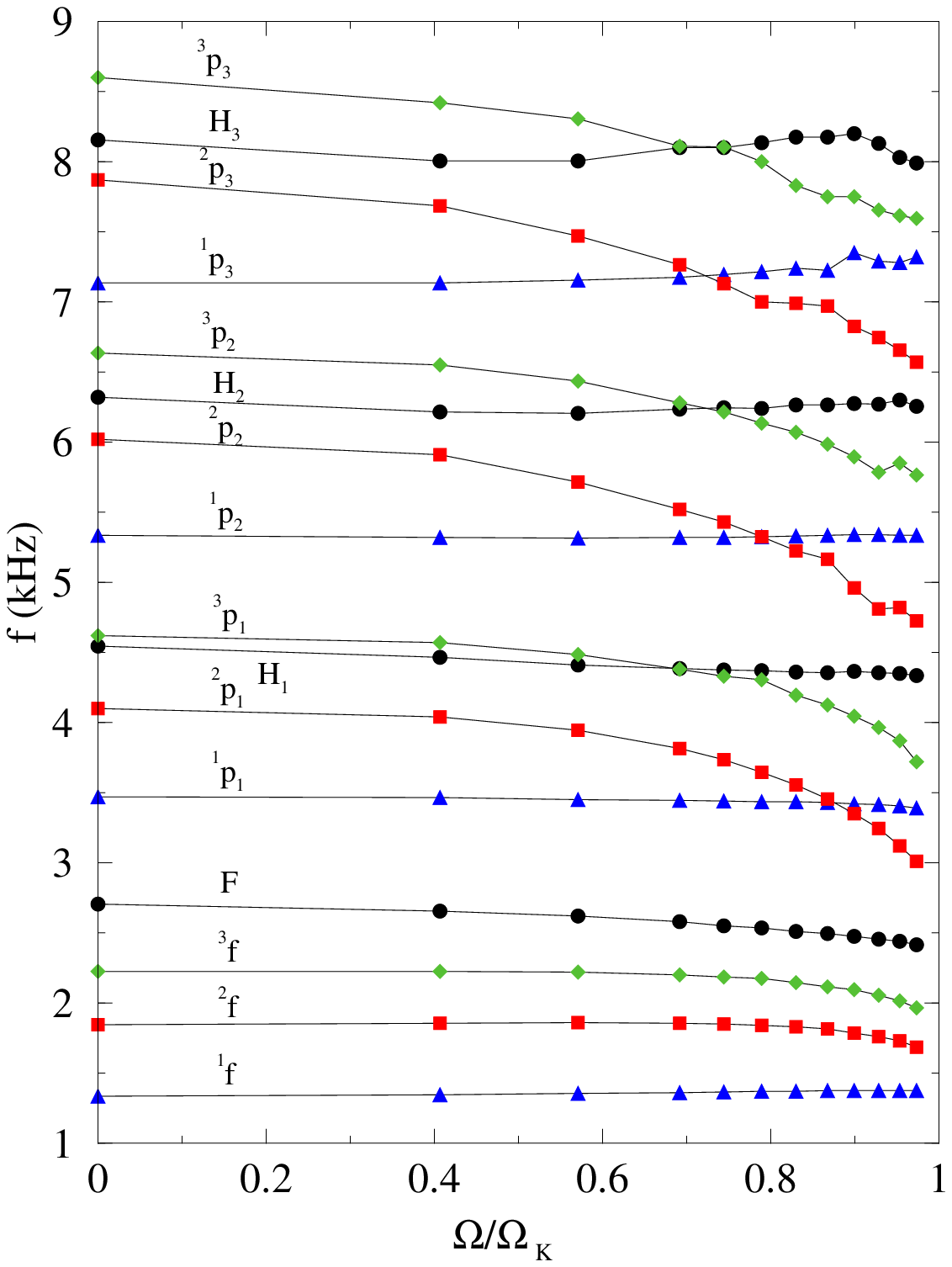}} \caption{\it
  Frequencies of several axisymmetric modes along a sequence of rapidly
  rotating relativistic polytropes of $N=1.0$, in the Cowling
  approximation.  On the horizontal axis, the angular velocity of each
  model is scaled to the angular velocity of the model at the
  mass-shedding limit. Lower-order modes are weakly affected by rapid
  rotation, while higher-order modes show apparent mode
  intersections. (Figure 10 of Font, Dimmelmeier, Gupta and
  Stergioulas, MNRAS \cite{Font01}.)}  \label{fig:axisym}
\end{figure}

Axisymmetric ($m=0$) pulsations in rotating relativistic stars could
be excited in a number of different astrophysical scenarios, such as
during core collapse, in star quakes induced by the secular spin-down
of a pulsar or during a large phase transition, in the merger of two
relativistic stars in a binary system etc. Due to rotational
couplings, the eigenfunction of any axisymmetric mode will involve a
sum of various spherical harmonics $Y_l^0$, so that even the
quasi-radial modes (with lowest-order $l=0$ contribution) would, in
principle, radiate gravitational waves.

Quasi-radial modes in rotating relativistic stars have been studied by
Hartle \& Friedman \cite{Hartle75} and by Datta et al. \cite{Datta98b}
in the slow-rotation approximation. Yoshida \& Eriguchi
\cite{Yoshida00} study quasi-radial modes of rapidly rotating stars in
the relativistic Cowling approximation and find that apparent
intersections between quasi-radial and other axisymmetric modes can
appear near the mass-shedding limit (see Figure \ref{fig_apparent}).
These apparent intersections are due to {\it avoided crossings}
between mode sequences, which are also known to occur for axisymmetric
modes of rotating Newtonian stars. Along a continuous sequence of
computed mode frequencies an avoided crossing occurs when another
sequence is encountered. In the region of the avoided crossing the
eigenfunctions of the two modes become of mixed character. Away from
the avoided crossing and along the continuous sequences of computed
mode frequencies, the eigenfunctions are exchanged. However, each
``quasi-normal mode'' is characterized by the shape of its
eigenfunction and thus, the sequences of computed frequencies that
belong to particular quasi-normal modes are discontinuous at avoided
crossings (see Figure \ref{fig_apparent} for more details). The
discontinuities can be found in numerical calculations, when
quasi-normal mode sequences are well resolved in the region of avoided
crossings. Otherwise, quasi-normal mode sequences will appear as
intersecting.

Several axisymmetric modes have recently been computed for rapidly
rotating relativistic stars in the Cowling approximation, using
time-evolutions of the nonlinear hydrodynamical equations
\cite{Font01} (see \cite{Font00} for a description of the 2D numerical
evolution scheme). As in \cite{Yoshida00}, Font et al. find that
apparent mode intersections are common for various higher-order
axisymmetric modes (see Figure \ref{fig:axisym}).  Axisymmetric inertial
modes also appear in the numerical evolutions.

The first fully relativistic frequencies of quasi-radial modes for
rapidly rotating stars (without assuming the Cowling approximation)
have been obtained recently, again through nonlinear time-evolutions
\cite{Font02} (see Section \ref{pulsrot}).

\begin{itemize} 
\item {\bf Going further} The stabilization of a relativistic star,
  that is marginally stable to axisymmetric perturbations, by an
  external gravitational field, is discussed in \cite{Th97}.
\end{itemize}  

\subsection{Nonaxisymmetric Perturbations}

\subsubsection{Nonrotating Limit}

Thorne, Campolattaro and Price, in a series of papers
\cite{TC67,PT69,Th69}, initiated the computation of nonradial modes by
formulating the problem in the Regge-Wheeler (RW) gauge \cite{RW} and
numerically computing nonradial modes for a number of neutron star
models. A variational method for obtaining eigenfrequencies and
eigenfunctions has been constructed by Detweiler and Ipser
\cite{DI73}. Lindblom and Detweiler \cite{LD83} explicitly reduced the
system of equations to four first-order ordinary differential
equations and obtained more accurate eigenfrequencies and damping
times for a larger set of neutron star models. They later realized
that their system of equations is sometimes singular inside the star
and obtained an improved set of equations which is free of this
singularity \cite{DL85}.

Chandrasekhar and Ferrari \cite{CF91} expressed the nonradial
pulsation problem in terms of a fifth-order system in a diagonal
gauge, which is formally independent of fluid variables. They thus
reformulate the problem in a way analogous to the scattering of
gravitational waves off a black hole. Ipser and Price \cite{IP91} show
that in the RW gauge, nonradial pulsations can be described by a
system of two second-order differential equations, which can also be
independent of fluid variables. In addition, they find that the
diagonal gauge of Chandrasekhar and Ferrari has a remaining gauge
freedom which, when removed, also leads to a fourth-order system of
equations \cite{PI91}.

In order to locate purely outgoing-wave modes, one has to be able to
distinguish the outgoing-wave part from the ingoing-wave part at
infinity. This is typically achieved using analytic approximations of
the solution at infinity.

$W$-modes pose a more challenging numerical problem because they are
strongly damped and the techniques used for $f-$ and $p-$modes fail to
distinguish the outgoing-wave part.  First accurate numerical solutions
were obtained by Kokkotas and Schutz \cite{Kokkotas92}, followed
by Leins, Nollert and Soffel \cite{Leins93}. Andersson, Kokkotas and Schutz
\cite{AKS95}, successfully combine a redefinition of variables with a
complex-coordinate integration method, obtaining highly accurate
complex frequencies for $w$ modes. In this method, the ingoing and
outgoing solutions are separated by numerically calculating their
analytic continuations to a place in the complex-coordinate plane,
where they have comparable amplitudes. Since this approach is purely
numerical, it could prove to be suitable for the computation of
quasi-normal modes in rotating stars, where analytic solutions at
infinity are not available.

The non-availability of asymptotic solutions at infinity in the case
of rotating stars is one of the major difficulties for computing
outgoing modes in rapidly rotating relativistic stars. A method that
may help to overcome this problem, at least to an acceptable
approximation, has been found by Lindblom, Mendell and Ipser
\cite{LMI97}.

The authors obtain approximate near-zone boundary conditions for the
outgoing modes that replace the outgoing-wave condition at infinity
and that enable one to compute the eigenfrequencies with very
satisfactory accuracy. First, the pulsation equations of polar modes
in the Regge-Wheeler gauge are reformulated as a set of two
second-order radial equations for two potentials - one corresponding
to fluid perturbations and the other to the perturbations of the
spacetime. The equation for the spacetime perturbation reduces to a
scalar wave equation at infinity and to Laplace's equation for
zero-frequency solutions. From these, an approximate boundary
condition for outgoing modes is constructed and imposed in the near
zone of the star (in fact on its surface) instead of at infinity.  For
polytropic models, the near-zone boundary condition yields $f$-mode
eigenfrequencies with real parts accurate to $0.01 \%-0.1 \%$ and
imaginary parts with accuracy at the $10 \% -20 \%$ level, for the
most relativistic stars. If the near zone boundary condition can be
applied to the oscillations of rapidly rotating stars, the resulting
frequencies and damping times should have comparable accuracy.

\subsubsection{Slow Rotation Approximation}

The slow rotation approximation is useful for obtaining a first
estimate of the effect of rotation on the pulsations of relativistic
stars.  To lowest order in rotation, a polar $l$-mode of an initially
nonrotating star couples to an axial $l \pm 1$ mode in the presence of
rotation.  Conversely, an axial $l$-mode couples to a polar $l \pm 1$
mode as was first discussed by Chandrasekhar and Ferrari \cite{CF91}.

The equations of nonaxisymmetric perturbations in the slow-rotation
limit are derived in a diagonal gauge, by Chandrasekhar and Ferrari 
\cite{CF91}, and in the Regge-Wheeler gauge,  by Kojima 
\cite{Koj92,Koj93}, where the complex frequencies $\sigma = \sigma_R + i \sigma_I$
for the $l=m$ modes of various polytropes are computed. For
counterrotating modes, both $\sigma_R$ and $\sigma_I$ decrease, tending to
zero, as the rotation rate increases (when $\sigma$ passes through zero,
the star becomes unstable to the CFS-instability).  Extrapolating
$\sigma_R$ and $\sigma_I$ to higher rotation rates, Kojima finds a large
discrepancy between the points where $\sigma_R$ and $\sigma_I$ go through
zero. This shows that the slow rotation formalism cannot accurately
determine the onset of the CFS-instability of polar modes in rapidly
rotating neutron stars.

In \cite{Koj93b}, it is shown that, for slowly rotating stars, the
coupling between polar and axial modes affects the frequency of $f$-
and $p$-modes only to second order in rotation, so that, in the slow
rotation approximation, to $O( \Omega)$, the coupling can be neglected
when computing frequencies.

The linear perturbation equations in the slow-rotation approximation
have recently been derived in a new gauge by Ruoff, Stavridis and
Kokkotas \cite{Ruoff02b}.  Using the ADM formalism, a first-order
hyperbolic evolution system is obtained, which is suitable for
numerical integration without further manipulations (as was required
in the Regge-Wheeler gauge). In this gauge (which is related to a
gauge introduced for nonrotating stars in \cite{Batiston71}) the
symmetry between the polar and axial equations becomes directly
apparent.

The case of relativistic inertial modes is different, as these modes
have both axial and polar parts at order $O(\Omega)$ and the presence of
continuous bands in the spectrum (at this order in the rotation rate)
has lead to a series of detailed investigations of the properties of
these modes (see \cite{Kokkotas02} for a review). In a recent paper,
Ruoff, Stavridis and Kokkotas \cite{Ruoff02} finally show that the
inclusion of both polar and axial parts in the computation of
relativistic $r$-modes, at order $O(\Omega)$, allows for discrete modes to
be computed, in agreement with post-Newtonian \cite{Lockitch01} and
nonlinear, rapid-rotation \cite{Stergioulas01} calculations.

\subsubsection{Post-Newtonian Approximation}

A step towards the solution of the perturbation equations in full
general relativity has been taken by Cutler and Lindblom
\cite{Cu91,CL92,Li95} who obtain frequencies for the $l=m$ $f$-modes
in rotating stars in the first post-Newtonian (1-PN) approximation.
The perturbation equations are derived in the post-Newtonian formalism
(see \cite{G91}), i.e. the equations are separated into equations of
consistent order in $1/c$.

Cutler and Lindblom show that in this scheme, the perturbation of the
1-PN correction of the four-velocity of the fluid can be obtained
analytically in terms of other variables; this is similar to  the perturbation 
in the three-velocity in the Newtonian Ipser-Managan scheme.  
The perturbation in the 1-PN corrections are
obtained by solving an eigenvalue problem, which consists of three
second order equations, with the 1-PN correction to the eigenfrequency
of a mode, $\Delta \omega$, as the eigenvalue.

Cutler and Lindblom obtain a formula that yields $\Delta \omega$ if one knows
the 1-PN stationary solution and the solution to the Newtonian
perturbation equations. Thus, the frequency of a mode in the 1-PN
approximation can be obtained without actually solving the 1-PN
perturbation equations numerically.  The 1-PN code was checked in the
nonrotating limit and it was found to reproduce the exact general
relativistic frequencies for stars with $M/R=0.2$, obeying an $N=1$
polytropic EOS, with an accuracy of $3 \% - 8 \%$.

Along a sequence of rotating stars, the frequency of a mode is
commonly described by the ratio of the frequency of the mode in the
comoving frame to the frequency of the mode in the nonrotating limit.
For an $N=1$ polytrope and for $M/R=0.2$, this frequency ratio is
reduced by as much as $12 \%$ in the 1-PN approximation compared to
its Newtonian counterpart (for the fundamental $l=m$ modes) which is
representative of the effect that general relativity has on the
frequency of quasi-normal modes in rotating stars.

\subsubsection{Cowling Approximation}

In several situations, the frequency of pulsations in relativistic
stars can be estimated even if one completely neglects the
perturbation in the gravitational field, i.e. if one sets $\delta
g_{ab}=0$ in the perturbation equations \cite{MVS83}.  In this
approximation, the pulsations are described only by the perturbation
in the fluid variables and the scheme works quite well for $f$, $p$
and $r$-modes \cite{LS90}. A different version of the Cowling
approximation, in which $\delta g_{tr}$ is kept nonzero in the
perturbation equations, has been suggested to be more suitable for
$g$-modes \cite{Fi88}, since these modes could have large fluid velocities,
even though the variation in the gravitational field is weak.

Yoshida and Kojima \cite{YK97}, examine the accuracy of the
relativistic Cowling approximation in slowly rotating stars.  The
first-order correction to the frequency of a mode depends only on the
eigenfrequency and eigenfunctions of the mode in the absence of
rotation and on the angular velocity of the star. The eigenfrequencies
of $f$, $p_1$ and $p_2$ modes for slowly rotating stars with $M/R$
between 0.05 and 0.2 are computed (assuming polytropic EOSs with $N=1$
and $N=1.5$) and compared to their counterparts in the slow- rotation
approximation.

For the $l=2$ $f$-mode, the relative error in the eigenfrequency
because of the Cowling approximation is $30 \%$ for weakly
relativistic stars ($M/R=0.05$) and about $15 \%$ for stars with
$M/R=0.2$ and the error decreases for higher $l$-modes. For the $p_1$
and $p_2$ modes the relative error is similar in magnitude but it is
smaller for less relativistic stars.  Also, for $p$-modes, the Cowling
approximation becomes more accurate for increasing radial mode number.

\begin{figure}[h]
  \def\epsfsize#1#2{0.4#1}
  \centerline{\epsfbox{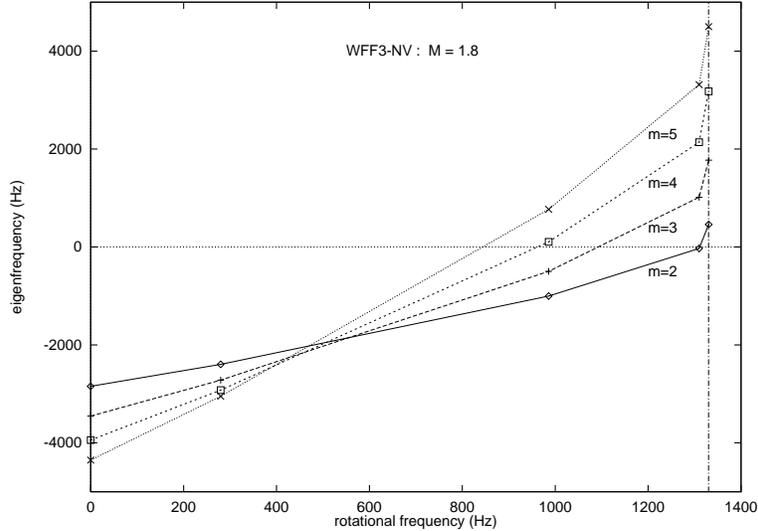}} \caption{\it
  Eigenfrequencies (in the Cowling approximation) of f-modes along a
  $M = 1.8 M_\odot$ sequence of models, constructed with the WFF3-NV EOS.
  The vertical line corresponds to the frequency of rotation of the
  model at the mass-shedding limit of the sequence. (Figure 1 of
  Yoshida and Eriguchi, ApJ \cite{Yoshida99}; used with permission.)}
\label{fig:cowling}
\end{figure}

As an application, Yoshida and Eriguchi \cite{YE97,Yoshida99} use the
Cowling approximation to estimate the onset of the $f$-mode CFS
instability in rapidly rotating relativistic stars and to compute
frequencies of $f$-modes for several realistic equations of state
 (see Figure \ref{fig:cowling}).

\subsection{Nonaxisymmetric Instabilities}

\subsubsection{Introduction}

     Rotating cold neutron stars, detected as pulsars, have a
     remarkably stable rotation period. But, at birth, or during
     accretion, rapidly rotating neutron stars can be subject to
     various nonaxisymmetric instabilities, which will affect the
     evolution of their rotation rate.
     
     If a protoneutron star has a sufficiently high rotation rate (so
     that e.g. $T/W > 0.27$ in the case of Maclaurin spheroids), it
     will be subject to a dynamical instability driven by
     hydrodynamics and gravity. Through the $l=2$ mode, the
     instability will deform the star into a bar shape.  This highly
     nonaxisymmetric configuration will emit strong gravitational
     waves with frequencies in the kHz regime. The development of the
     instability and the resulting waveform have been computed
     numerically in the context of Newtonian gravity by Houser et al.
     \cite{HCS94} and in full general relativity by Shibata et al.
     \cite{Shibata00c} (see Section \ref{s:dynamical}).
     
     At lower rotation rates, the star can become unstable to secular
     nonaxisymmetric instabilities, driven by gravitational radiation
     or viscosity. Gravitational radiation drives a nonaxisymmetric
     instability when a mode that is retrograde in a frame corotating
     with the star appears as prograde to a distant inertial observer,
     via the Chandrasekhar-Friedman-Schutz (CFS) mechanism
     \cite{C70,FS78}: A mode that is retrograde in the corotating
     frame has negative angular momentum, because the perturbed star
     has less angular momentum than the unperturbed one. If, for a
     distant observer, the mode is prograde, it removes positive
     angular momentum from the star and thus the angular momentum of
     the mode becomes increasingly negative.
     
     The instability evolves on a secular timescale, during which the
     star loses angular momentum via the emitted gravitational waves.
     When the star rotates slower than a critical value, the mode
     becomes stable and the instability proceeds on the longer
     timescale of the next unstable mode, unless it is suppressed by
     viscosity.
     
     Neglecting viscosity, the CFS-instability is generic in rotating
     stars for both polar and axial modes.  For polar modes, the
     instability occurs only above some critical angular velocity,
     where the frequency of the mode goes through zero in the inertial
     frame.  The critical angular velocity is smaller for increasing
     mode number $l$. Thus, there will always be a high enough mode
     number $l$, for which a slowly rotating star will be unstable.
     Many of the hybrid inertial modes (and in particular the
     relativistic $r$-mode) are generically unstable in all rotating
     stars, since the mode has zero frequency in the inertial frame
     when the star is nonrotating \cite{A97,FM97}.
     
     The shear and bulk viscosity of neutron star matter is able to
     suppress the growth of the CFS-instability except when the star
     passes through a certain temperature window. In Newtonian
     gravity, it appears that the polar mode CFS-instability can occur
     only in nascent neutron stars that rotate close to the
     mass-shedding limit \cite{IL91a,IL91b,IL92,YE95,LM95}, but the
     computation of neutral $f$-modes in full relativity
     \cite{SPHD,SF97} shows that relativity enhances the instability,
     allowing it to occur in stars with smaller rotation rates than
     previously thought.

\begin{itemize}
\item {\bf Going further.} A numerical method for the analysis of the
  ergosphere instability in relativistic stars, which could be
  extended to nonaxisymmetric instabilities of fluid modes, is
  presented by Yoshida and Eriguchi in \cite{YE96}.
\end{itemize}

\subsubsection{CFS-Instability of Polar Modes}

The existence of the CFS-instability in rotating stars was first
demonstrated by Chandrasekhar \cite{C70} in the case of the $l=2$ mode
in uniformly rotating, uniform density Maclaurin spheroids. Friedman
and Schutz \cite{FS78} show that this instability also appears in
compressible stars and that all rotating self-gravitating perfect
fluid configurations are generically unstable to the emission of
gravitational waves. In addition, they find that a nonaxisymmetric
mode becomes unstable when its frequency vanishes in the inertial
frame. Thus, zero-frequency outgoing-modes in rotating stars are
neutral (marginally stable).

In the Newtonian limit, neutral modes have been determined for several
polytropic EOSs \cite{IFD85,M85,IL90,YE95}. The instability first sets
in through $l=m$ modes. Modes with larger $l$ become unstable at lower
rotation rates but viscosity limits the interesting ones to $l \leq5$.
For an $N=1$ polytrope, the critical values of $T/W$ for the $l=3,4$
and 5 modes are 0.079, 0.058 and 0.045 respectively and these values
become smaller for softer polytropes.  The $l=m=2$ "bar" mode has a
critical $T/W$ ratio of 0.14 that is almost independent of the
polytropic index. Since soft EOSs cannot produce models with high
$T/W$ values, the bar mode instability appears only for stiff
Newtonian polytropes of $N \leq 0.808$ \cite{Ja64,SL96}.  In addition,
the viscosity-driven bar mode appears at the same critical $T/W$ ratio
as the bar mode driven by gravitational radiation \cite{IM85} (we will
see later that this is no longer true in general relativity).

\begin{figure}[p]
  \def\epsfsize#1#2{0.7#1} \centerline{\epsfbox{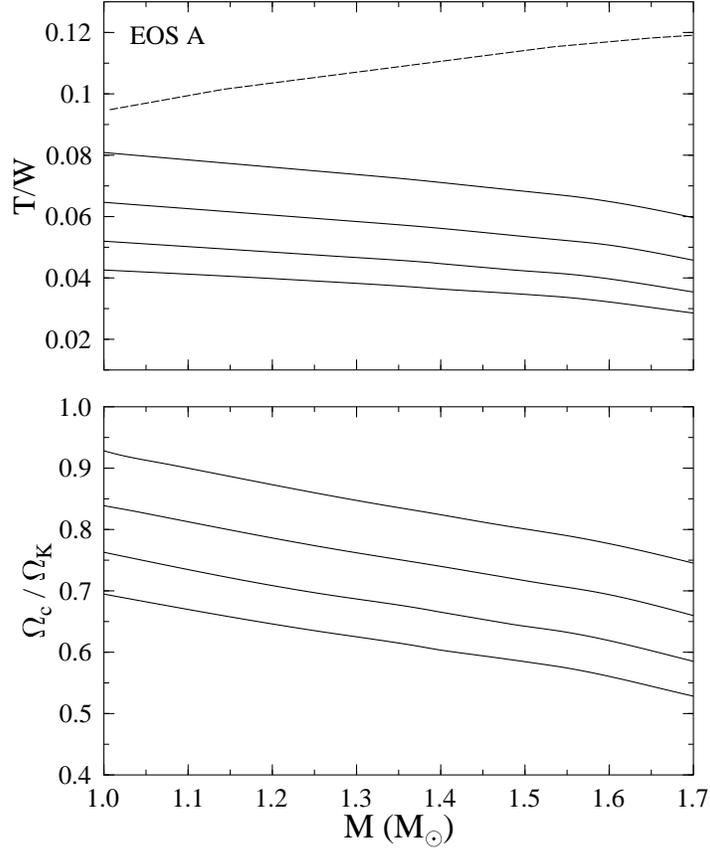}}
  \caption{\it The $l=m$ neutral $f$-mode sequences for EOS A. Shown
  are the ratio of rotational to gravitational energy $T/W$ (upper
  panel) and the ratio of the critical angular velocity $\Omega_c$ to the
  angular velocity at the mass-shedding limit for uniform rotation
  (lower panel) as a function of gravitational mass. The solid curves
  are the neutral mode sequences for $l=m=2, 3, 4$ and 5 (from top to
  bottom), while the dashed curve in the upper panel corresponds to
  the mass-shedding limit for uniform rotation. The $l=m=2$ $f$-mode
  becomes CFS-unstable even at $85\%$ of the mass-shedding limit, for
  $1.4$ $M_\odot$ models constructed with this EOS. (Figure 2 of Morsink,
  Stergioulas and Blattning, ApJ \cite{MSB98}.)}  \label{fig:msb}
\end{figure}

\begin{figure}[p]
  \def\epsfsize#1#2{0.6#1} \centerline{\epsfbox{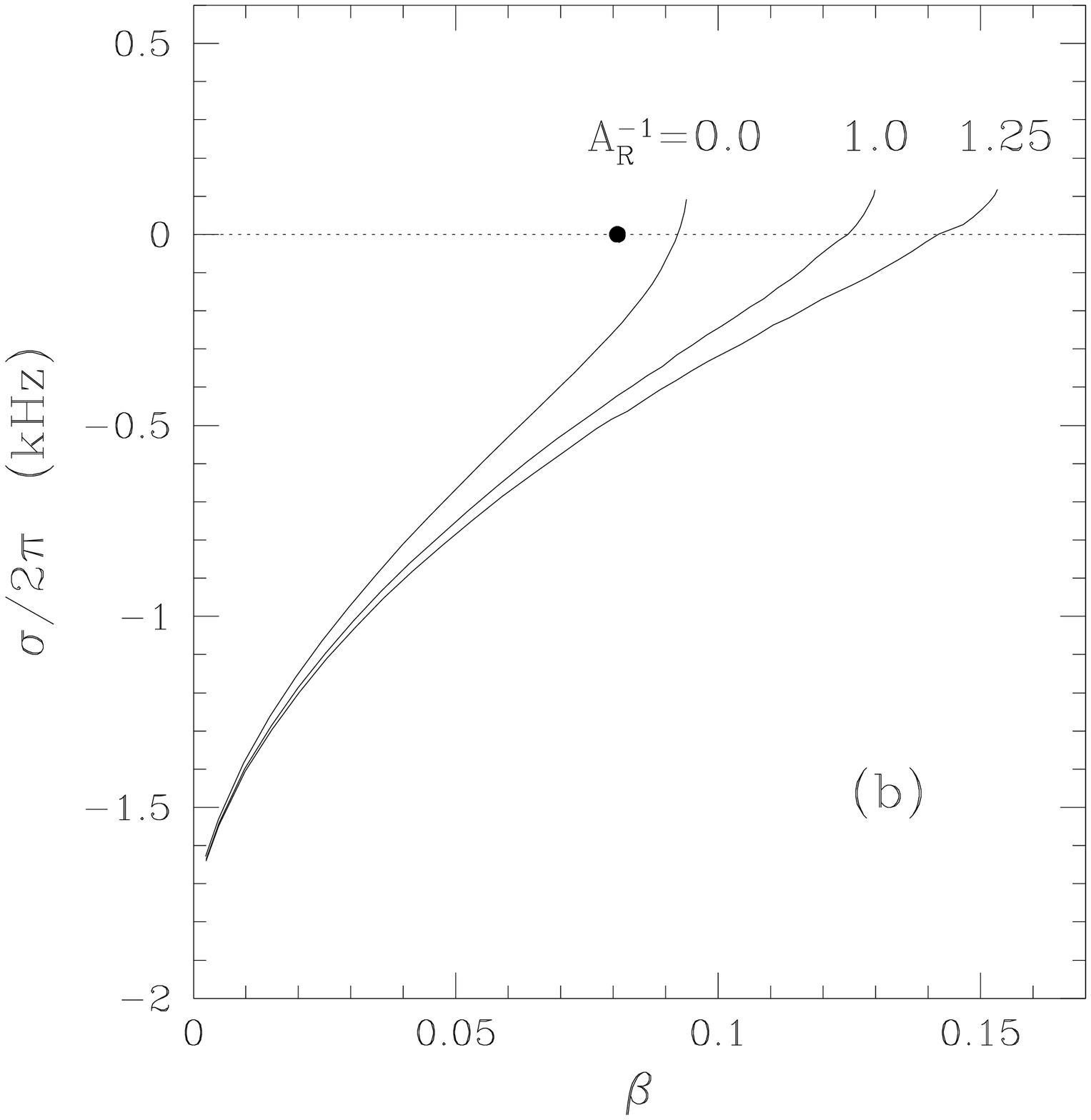}}
  \caption{\it Eigenfrequencies (in the Cowling approximation) of the
  $m=2$ mode as a function of the parameter $\beta=T/|W|$ for three
  different sequences of differentially rotating neutron stars (the
  $A_{_{\rm R}}^{-1}=0.0$ line corresponding to uniform rotation). The
  filled dot indicates the neutral stability point of a uniformly
  rotating star computed in full General Relativity (Stergioulas and
  Friedman, 1998 \cite{SF97}). Differential rotation shifts the
  neutral point to higher rotation rates. (Figure 1 of Yoshida,
  Rezzolla, Karino and Eriguchi in ApJ \cite{Yoshida02}; used with
  permission.)}  \label{fig:diff}
\end{figure}

The post-Newtonian computation of neutral modes by Cutler and Lindblom
\cite{CL92,Li95} has shown that general relativity tends to strengthen
the CFS-instability. Compared to their Newtonian counterparts,
critical angular velocity ratios $\Omega_c/\Omega_0$ (where
$\Omega_0=(3M_0/4R_0^3)^{1/2}$ and $M_0$, $R_0$ are the mass and radius of
the nonrotating star in the sequence), are lowered by as much as $10
\%$ for stars obeying the $N=1$ polytropic EOS (for which the
instability occurs only for $l=m \geq 3$ modes in the post-Newtonian
approximation).

In full general relativity, neutral modes have been determined for
polytropic EOSs of $N \geq 1.0$ by Stergioulas and Friedman
\cite{SPHD,SF97}, using a new numerical scheme. The scheme completes
the Eulerian formalism developed by Ipser \& Lindblom in the Cowling
approximation (where $\delta g_{ab}$ was neglected) \cite{IL92}, by
finding an appropriate gauge in which the time-independent
perturbation equations can be solved numerically for $\delta g_{ab}$.  The
computation of neutral modes for polytropes of $N=1.0$, $1.5$ and
$2.0$ shows that relativity significantly strengthens the instability.
For the $N=1.0$ polytrope, the critical angular velocity ratio $\Omega_c /
\Omega_K$, where $\Omega_K$ is the angular velocity at the mass-shedding limit
at same central energy density, is reduced by as much as $15 \%$ for
the most relativistic configuration (see Figure \ref{fig:msb}).  A
surprising result (which was not found in computations that used the
post-Newtonian approximation), is that the $l=m=2$ bar mode is
unstable even for relativistic polytropes of index $N=1.0$. The
classical Newtonian result for the onset of the bar mode instability
($N_{crit} <0.808$) is replaced by
\begin{equation}
                 N_{crit} <1.3,
\end{equation}
in general relativity.  It is evident that, in relativistic stars, the
onset of the gravitational-radiation-driven bar mode does not coincide
with the onset of the viscosity-driven bar mode, which occurs at
larger $T/W$ \cite{BFG97}.  The computation of the onset of the
CFS-instability in the relativistic Cowling approximation by Yoshida
and Eriguchi \cite{YE97} agrees qualitatively with the conclusions in
\cite{SPHD,SF97}.

Morsink, Stergioulas and Blattning \cite{MSB98} extend the method
presented in \cite{SF97} to a wide range of realistic equations of
state (which usually have a stiff high density region, corresponding
to polytropes of index $N=0.5-0.7$) and find that the $l=m=2$ bar mode
becomes unstable for stars with gravitational mass as low as $1.0 -
1.2M_\odot$.  For $1.4M_\odot$ neutron stars, the mode becomes unstable at
$80 \% -95 \%$ of the maximum allowed rotation rate.  For a wide range
of equations of state, the $l=m=2$ $f$-mode becomes unstable at a
ratio of rotational to gravitational energies $T/W \sim 0.08$ for $1.4
M_\odot$ stars and $T/W \sim 0.06$ for maximum mass stars.  This is to be
contrasted with the Newtonian value of $T/W \sim 0.14$.  The empirical
formula
\begin{equation}
  \left( T/W \right)_2 = 0.115 -0.048 \frac{M}{M_{\rm max}^{\rm sph}}, \label{emp}
\end{equation}
where $M_{\rm max}^{\rm sph}$ is the maximum mass for a spherical star
allowed by a given equation of state, gives the critical value of
$T/W$ for the bar $f-$mode instability, with an accuracy of $4\%-6$\%,
for a wide range of realistic EOSs.

 In newly-born neutron stars the CFS-instability could develop while
 the background equilibrium star is still differentially rotating. In
 that case, the critical value of $T/W$, required for the instability
 in the $f$-mode to set in, is larger than the corresponding value in
 the case of uniform rotation \cite{Yoshida02} (Figure \ref{fig:diff}).
 The mass-shedding limit for differentially rotating stars also
 appears at considerably larger $T/W$ than the mass-shedding limit for
 uniform rotation. Thus, Yoshida et al.  \cite{Yoshida02} suggest that
 differential rotation favours the instability, since the ratio
 $(T/W)_{\rm critical}/(T/W)_{\rm shedding}$ decreases with increasing
 degree of differential rotation.

\subsubsection{CFS-Instability of Axial Modes}
\label{s_axial}
     
     In nonrotating stars, axial fluid modes are degenerate at
zero frequency but in rotating stars they have nonzero frequency and
are called $r$-modes in the Newtonian limit \cite{PP78,Sa82}. To 
order $O(\Omega)$, their frequency in the inertial frame is
\begin{equation}
     \omega_i = -m\Omega \Bigl(1-\frac{2}{l(l+1)} \Bigr), \label{e:om}
\end{equation}
while the radial eigenfunction of the perturbation in the velocity can
be determined at order $\Omega^2$ \cite{Koj97}. According to (\ref{e:om}),
$r$-modes with $m>0$ are prograde ($\omega_i<0$) with respect to a distant
observer but retrograde ($\omega_r = \omega_i+m\Omega >0$) in the comoving frame
for all values of the angular velocity.  Thus, $r$-modes in
relativistic stars are generically unstable to the emission of
gravitational waves via the CFS-instability, which was first
discovered by Andersson \cite{A97}, in the case of slowly rotating,
relativistic stars.  This result was proved rigorously by Friedman and
Morsink \cite{FM97}, who showed that the canonical energy of the modes
is negative.

\begin{figure}[t]
  \def\epsfsize#1#2{0.5#1} \centerline{\epsfbox{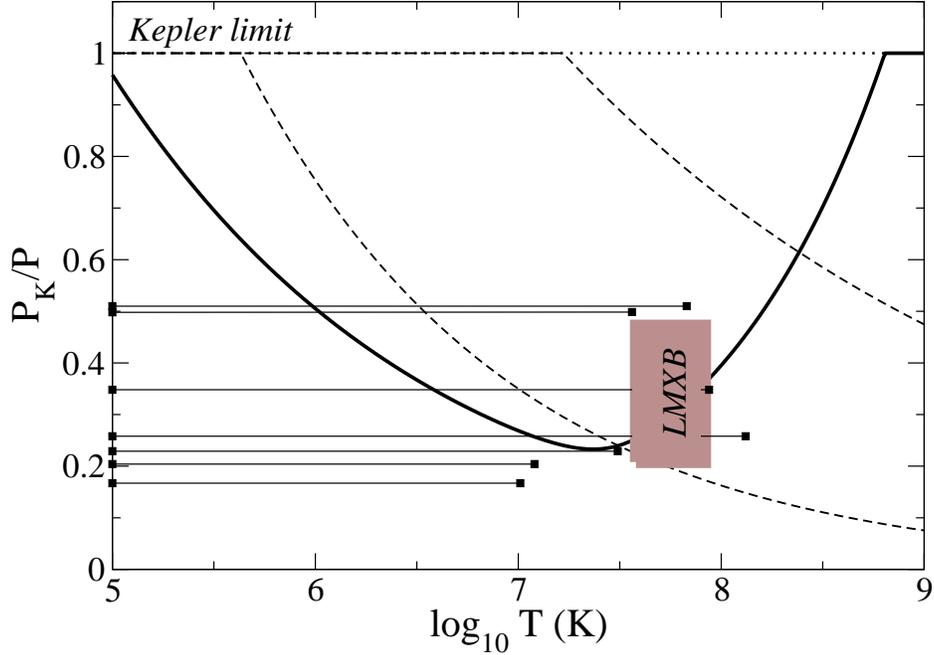}}
  \caption{\it The r-mode instability window for a strange star of
  $M=1.4M_\odot$ and $R=10$~km (solid line). Dashed curves show the
  corresponding instability windows for normal npe fluid and neutron
  stars with a crust. The instability window is compared to i) the
  inferred spin-periods for accreting stars in Low-mass X-ray binaries
  [shaded box], and ii) the fastest known millisecond pulsars (for
  which observational upper limits on the temperature are available)
  [horizontal lines]. (Figure 1 of Andersson, Jones and Kokkotas
  \cite{Andersson02c}; used with permission.)} \label{fig:rstrange}
\end{figure}

\begin{figure}[h]
  \def\epsfsize#1#2{0.4#1}
  \centerline{\epsfbox{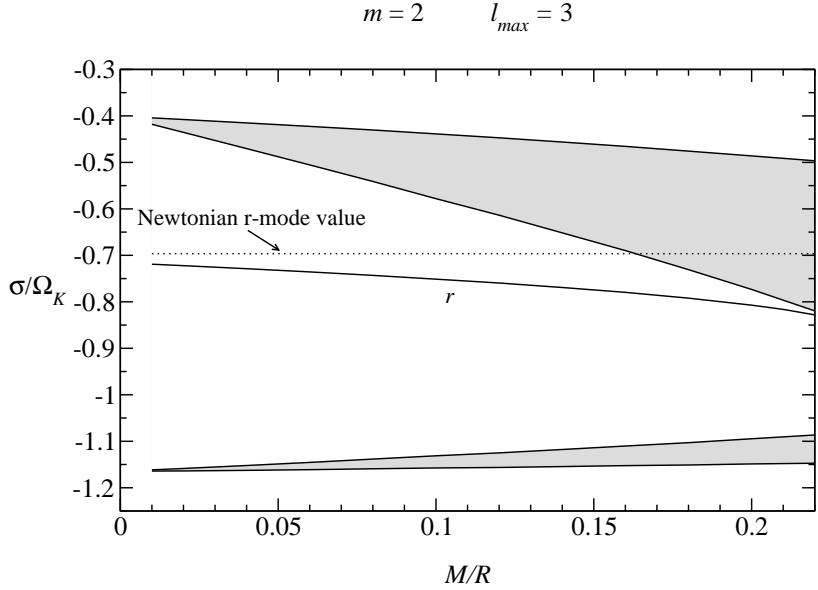}} \caption{\it
  Relativistic $r$-mode frequencies for a range of the compactness
  ratio $M/R$. The coupling of polar and axial terms, even in the
  order $O(\Omega)$ slow-rotation approximation has a dramatic impact on
  the continuous frequency bands (shaded areas), allowing the $r$-mode
  to exist even in highly compact stars.  The Newtonian value of the
  $r$-mode frequency is plotted as a dashed-dotted line. (Figure 3 of
  Ruoff, Stavridis and Kokkotas \cite{Ruoff02}; used with
  permission.)}  \label{fig:rspect}
\end{figure}

\begin{figure}[p]
  \def\epsfsize#1#2{0.6#1} \centerline{\epsfbox{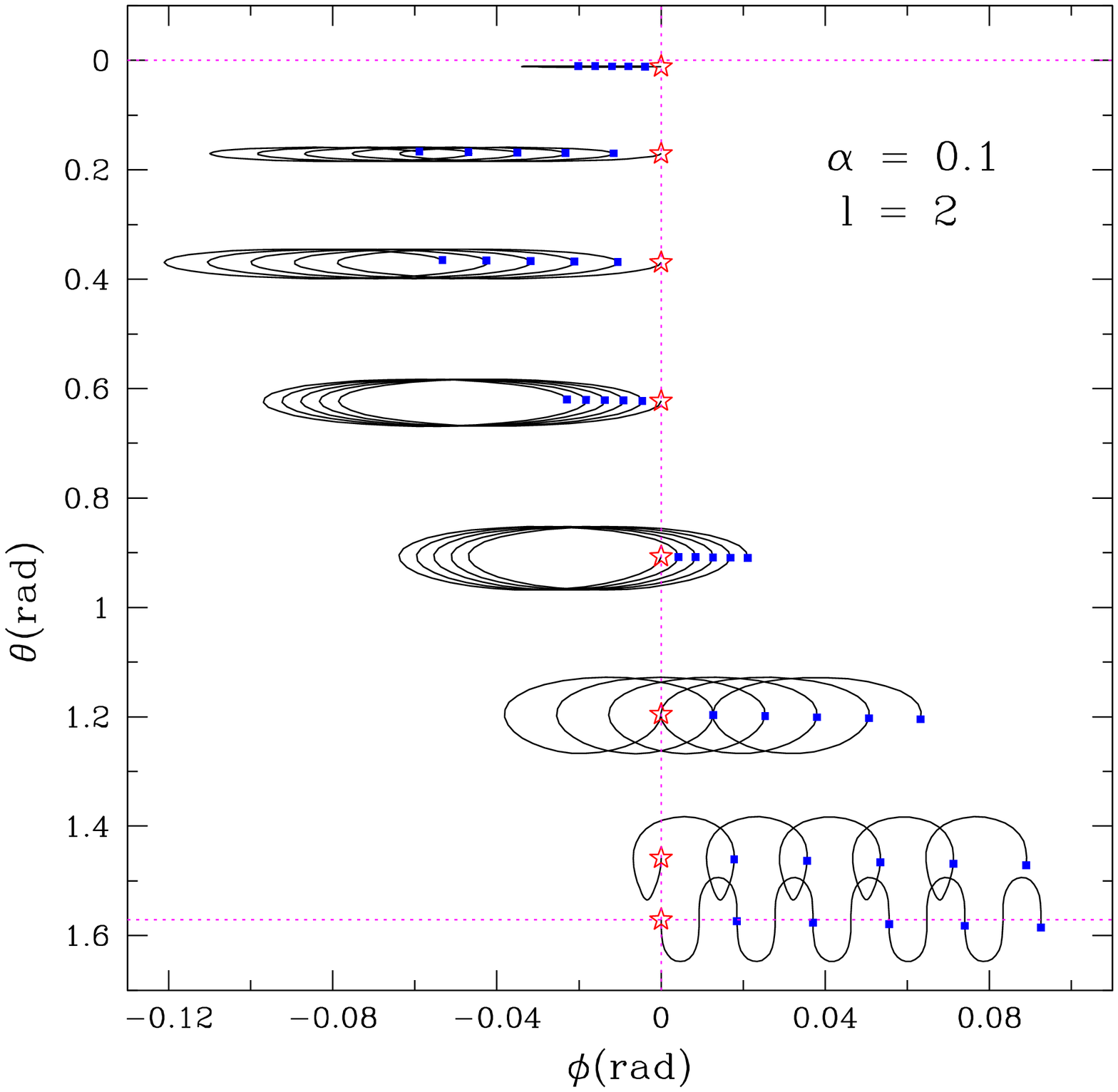}}
  \caption{\it Projected trajectories of several fiducial fluid
  elements (as seen in the corotating frame) for an $l=m=2$ Newtonian
  $r$-mode.  All of the fluid elements are initially positioned on the
  $\phi_0=0$ meridian, at different latitudes (indicated with
  stars). Blue dots indicate the position of the fluid elements after
  each full oscillation period. The $r$-mode induces a kinematical,
  differential drift. (Figure 2c of Rezzolla, Lamb, Markovi{\'c} and
  Shapiro, PRD \cite{Rezzolla01b}; used with permission.)}
  \label{fig:rdiff}
\end{figure}

Two independent computations in the Newtonian Cowling approximation
\cite{LOM98,AKS98} showed that the usual shear and bulk viscosity
assumed to exist for neutron star matter is not able to damp the
$r$-mode instability, even in slowly rotating stars. In a temperature
window of $10^5$ K $<T< 10^{10}$ K, the growth time of the $l=m=2$
mode becomes shorter than the shear or bulk viscosity damping time at
a critical rotation rate that is roughly one tenth the maximum allowed
angular velocity of uniformly rotating stars. The gravitational
radiation is dominated by the mass current quadrupole term.  These results
suggested that a rapidly rotating proto-neutron star will spin down to
Crab-like rotation rates within one year of its birth, because of the
$r$-mode instability.  Due to uncertainties in the actual viscosity
damping times and because of other dissipative mechanisms, this
scenario also is consistent with somewhat higher initial spins, like
the suggested initial spin of a several milliseconds of the X-ray
pulsar in the supernova remnant N157B \cite{Ma98}.  Millisecond
pulsars with periods less than a few milliseconds can then only form after the
accretion-induced spin-up of old pulsars and not in the
accretion-induced collapse of a white dwarf.

The precise limit on the angular velocity of newly-born neutron stars
will depend on several factors, such as the strength of the bulk
viscosity, the cooling process, superfluidity, presence of hyperons,
influence of a solid crust etc.  In the uniform density approximation,
the $r$-mode instability can be studied analytically to $O(\Omega^2)$ in
the angular velocity of the star \cite{KS98}. A study on the
issue of detectability of gravitational waves from the $r$-mode
instability was presented in \cite{OW98} (see section \ref{grw}),
while Andersson, Kokkotas and Stergioulas \cite{AKSt98} and 
Bildsten \cite{Bildsten98} proposed that
the $r$-mode instability is limiting the spin of millisecond pulsars
spun-up in LMXBs and it could even set the minimum observed spin period of
$\sim 1.5$ ms (see \cite{Andersson00}). This scenario is also compatible
with observational data, if one considers strange stars instead of
neutron stars \cite{Andersson02c} (see Figure \ref{fig:rstrange}).

Since the discovery of the $r$-mode instability, a large number of
authors have studied in more detail the development of the
instability and its astrophysical consequences. Unlike in the case of
the $f$-mode instability, many different aspects and interactions have
been considered. This intense focus on the detailed physics has been
very fruitful and we now have a much more complete understanding of
the various physical processes that are associated with pulsations in
rapidly rotating relativistic stars. The latest understanding of the
$r$-mode instability is that it may not be a very promising
gravitational wave source (as originally thought), but the important
astrophysical consequences, such as the limits of the spin of young
and of recycled neutron stars are still considered plausible.  The
most crucial factors affecting the instability are magnetic fields
\cite{Spruit99,Rezzolla00,Rezzolla01b,Rezzolla01c}, 
possible hyperon bulk viscosity
\cite{JonesPB01,Lindblom01,Haensel02} and nonlinear saturation
\cite{Stergioulas01,Lindblom01b,Lindblom02,Arras02}. The question of
the possible existence of a continuous spectrum has also been
discussed by several authors, but the most recent analysis suggests
that higher-order rotational effects still allow for discrete $r$-modes in
relativistic stars \cite{Yoshida02b,Ruoff02} (see Figure \ref{fig:rspect}).

\begin{figure}[t]
  \def\epsfsize#1#2{0.6#1} \centerline{\epsfbox{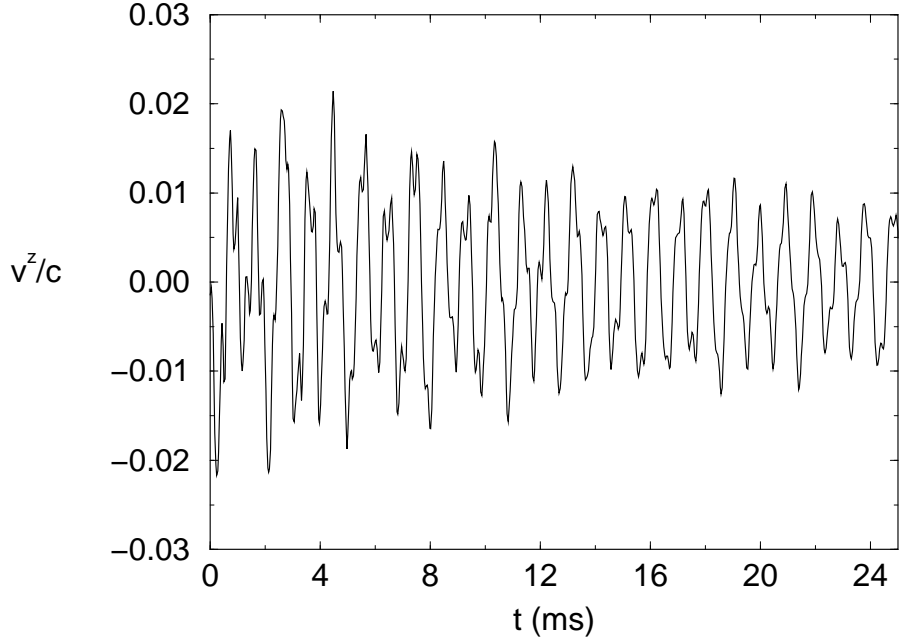}}
  \caption{\it Evolution of the axial velocity in the equatorial plane
  for a relativistic $r$-mode in a rapidly rotating $N=1.0$ polytrope
  (in the Cowling approximation). Since the initial data used to
  excite the mode are not exact, the evolution is a superposition of
  (mainly) the $l=m=2$ $r$-mode and several inertial modes. The
  amplitude of the oscillation decreases due to numerical
  (finite-differencing) viscosity of the code. A beating between the
  $l=m=2$ $r$-mode and another inertial mode can also be seen. (Figure
  2 of Stergioulas and Font, PRL \cite{Stergioulas01}.)}
  \label{fig:rmode}
\end{figure}

Magnetic fields can affect the $r$-mode instability, as the $r$-mode
velocity field creates differential rotation, which is both
kinematical and due to gravitational radiation reaction (see Figure
\ref{fig:rdiff}). Under differential rotation, an initially weak
poloidal magnetic field is wound-up, creating a strong toroidal field,
which causes the $r$-mode amplitude to saturate. If neutron stars have
hyperons in their cores, the associated bulk viscosity is so strong
that it could completely prevent the growth of the $r$-mode
instability. However, hyperons are predicted only by certain equations
of state and the relativistic mean field theory is not universally
accepted. Thus, our ignorance of the true equation of state still
leaves a lot of room for the $r$-mode instability to be considered
viable.

The gravitational-wave detection from $r$-modes depends crucially on
the nonlinear saturation amplitude. A first study by Stergioulas \&
Font \cite{Stergioulas01} suggests that $r$-modes can exist at large
amplitudes of order unity for dozens of rotational periods in rapidly
rotating relativistic stars (Figure \ref{fig:rmode}).  The study used
3D relativistic hydrodynamical evolutions in the Cowling
approximation.  This result was confirmed by Newtonian 3D simulations
of nonlinear $r$-modes by Lindblom, Tohline \& Vallisneri
\cite{Lindblom01,Lindblom01b}. Lindblom et al. went further, using an
accelerated radiation reaction force, to artificially grow the
$r$-mode amplitude on a hydrodynamical (instead of the secular)
timescale. At the end of the simulations, the $r$-mode grew so large
that large shock waves appeared on the surface of the star, while the
amplitude of the mode subsequently collapsed.  Lindblom et
al. suggested that shock heating may be the mechanism that saturates
the $r$-modes at a dimensionless amplitude of $\alpha \sim 3$.  

More recent studies of nonlinear couplings between the $r$-mode and
higher-order inertial modes \cite{Arras02} and new 3D nonlinear
Newtonian simulations \cite{Gressman02} seem to suggest a different
picture. The $r$-mode could be saturated due to mode couplings or due
to a hydrodynamical instability at amplitudes much smaller than the
amplitude at which shock waves appeared in the simulations by Lindblom
et al. Such a low amplitude, on the other hand, modifies the
properties of the $r$-mode instability as a gravitational wave source,
but is not necessarily bad news for gravitational wave detection, as a
lower spin-down rate also implies a higher event rate for the $r$-mode
instability in LMXBs in our own Galaxy \cite{Andersson02c,Heyl02}.
The 3D simulations need to achieve significantly higher resolutions
before definite conclusions can be reached, while the Arras et
al. work could be extended to rapidly rotating relativistic stars (in
which case the mode frequencies and eigenfunctions could change
significantly, compared to the slowly rotating Newtonian case, which
could affect the nonlinear coupling coefficients). Spectral methods
can be used for achieving high accuracy in mode calculations and first
results have been obtained by Villain and Bonazzolla \cite{Villain02}
for inertial modes of slowly rotating stars in the relativistic
Cowling approximation.

For a more extensive coverage of the numerous articles on the $r$-mode
instability that appeared in recent years, the reader is referred to
several excellent recent review articles
\cite{Andersson01c,Friedman01,Lindblom01c,Kokkotas02,Andersson02b}.

\begin{itemize}
  \item {\bf Going further.} If rotating stars with very high compactness exist,
then $w$-modes can also become unstable, as was recently found by Kokkotas,
Ruoff and Andersson \cite{Kokkotas03}. The possible astrophysical implications
are still under investigation.
\end{itemize}

\subsubsection{Effect of Viscosity on CFS-Instability}

     In the previous sections, we have discussed the growth of the 
CFS-instability driven by gravitational radiation in an 
otherwise nondissipative
star. The effect of neutron star matter viscosity on the dynamical
evolution of nonaxisymmetric perturbations can be considered
separately, when the timescale of the viscosity is much longer than
the oscillation timescale. If $\tau_{GR}$ is the computed growth rate of
the instability in the absence of viscosity and $\tau_s$, $\tau_b$ are
the timescales of shear and bulk viscosity, then the total timescale
of the perturbation is
\begin{equation}
      \frac{1}{\tau} = \frac{1}{\tau_{GR}} + \frac{1}{\tau_s} + 
                       \frac{1}{\tau_b}.
\end{equation}
Since $\tau_{GR} <0$ and $\tau_b$, $\tau_s>0$, a mode will grow only if
$\tau_{GR}$ is shorter than the viscous timescales, so that $1/\tau<0$.

In normal neutron star matter, shear viscosity is dominated by
neutron-neutron scattering with a temperature dependence of $T^{-2}$
\cite{FI76} and computations in the Newtonian limit and post-Newtonian
approximation show that the CFS-instability is suppressed for $T
<10^6$ K - $10^7$ K \cite{IL91a, IL91b, YE95, Li95}.  If neutrons
become a superfluid below a transition temperature $T_{s}$, then
mutual friction, which is caused by the scattering of electrons off
the cores of neutron vortices could significantly suppress the $f$-mode
instability for $T<T_{s}$ \cite{Lindblom95} but the
$r$-mode instability remains unaffected \cite{Lindblom00}.
The superfluid transition temperature depends on the theoretical model
for superfluidity and lies in the range $10^8$ K - $6 \times 10^9$ K
\cite{Pa94}.
   
In a pulsating fluid that undergoes compression and expansion, the
weak interaction requires a relatively long time to re-establish
equilibrium.  This creates a phase lag between density and pressure
perturbations, which results in a large bulk viscosity \cite{Sa89}.
The bulk viscosity due to this effect can suppress the CFS-instability
only for temperatures for which matter has become transparent to
neutrinos \cite{LS95,BFG96}.  It has been proposed that for
$T>5 \times 10^9$K, matter will be opaque to neutrinos and the neutrino
phase space could be blocked (\cite{LS95} see also \cite{BFG96}).  In
this case, bulk viscosity will be too weak to suppress the
instability, but a more detailed study is needed.

In the neutrino transparent regime, the effect of bulk viscosity on
the instability depends crucially on the proton fraction $x_p$. If
$x_p$ is lower than a critical value ($\sim \frac{1}{9}$), only modified
URCA processes are allowed and bulk viscosity limits, but does not
suppress completely, the instability \cite{IL91a, IL91b, YE95}. For
most modern EOSs, however, the proton fraction is larger than $\sim
\frac{1}{9}$ at sufficiently high densities \cite{Lat91}, allowing
direct URCA processes to take place. In this case, depending on the
EOS and the central density of the star, the bulk viscosity could
almost completely suppress the CFS-instability in the neutrino
transparent regime \cite{Zd95}. At high temperatures, $T>5 \times 10^9$ K,
even if the star is opaque to neutrinos, the direct URCA cooling timescale
to $T~5 \times 10^9$ K could be shorter than the the growth timescale of
the CFS instability.

\subsubsection{Gravitational Radiation from CFS-Instability}
\label{grw}

Conservation of angular momentum and the inferred initial period
(assuming magnetic braking) of a few milliseconds for the X-ray pulsar
in the supernova remnant N157B \cite{Ma98}, suggests that a fraction
of neutron stars may be born with very large rotational energies. The
$f$-mode bar CFS-instability thus appears as a promising source for
the planned gravitational wave detectors \cite{LS95}. It could also
play a role in the rotational evolution of merged binary neutron
stars, if the post-merger angular momentum exceeds the maximum allowed
to form a Kerr black hole \cite{BaS98} or if differential rotation
temporarily stabilizes the merged object.

Lai and Shapiro \cite{LS95} have studied the development of the
$f$-mode instability using Newtonian ellipsoidal models
\cite{LRS93,LRS94}. They consider the case when a rapidly rotating
neutron star is created in a core collapse. After a brief dynamical
phase, the protoneutron star becomes secularly unstable. The
instability deforms the star into a nonaxisymmetric configuration via
the $l=2$ bar mode. Since the star loses angular momentum via the
emission of gravitational waves, it spins-down until it becomes
secularly stable.  The frequency of the waves sweeps downward from a
few hundred Hz to zero, passing through LIGO's ideal sensitivity band.
A rough estimate of the wave amplitude shows that, at $\sim100$Hz, the
gravitational waves from the CFS-instability could be detected out to
the distance of 140Mpc by the advanced LIGO detector.  This result is
very promising, especially since for relativistic stars the
instability will be stronger than the  Newtonian estimate
\cite{SF97}. Whether $r$-modes should also be considered a promising
gravitational-wave source depends crucially on their nonlinear saturation
amplitude (see Section \ref{s_axial}).

\begin{itemize}
  \item {\bf Going further.} The possible ways for neutron stars
   to emit gravitational waves and their detectability are reviewed in 
   \cite{BG96,BGou,GBG96,FL98,Th96,Sc98,Cutler02}. 
\end{itemize}

\subsubsection{Viscosity-Driven Instability}

A different type of nonaxisymmetric instability in rotating stars is
the instability driven by viscosity, which breaks the circulation of
the fluid \cite{RS63,Ja64}. The instability is suppressed by
gravitational radiation, so it cannot act in the temperature window in
which the CFS-instability is active. The instability
sets in when the frequency of an $l=-m$ mode goes through zero in the
rotating frame. In contrast to the CFS-instability, the
viscosity-driven instability is not generic in rotating stars. The
$m=2$ mode becomes unstable at a high rotation rate for very stiff
stars and higher $m$-modes become unstable at larger rotation rates.

In Newtonian polytropes, the instability occurs only for stiff
polytropes of index $N<0.808$ \cite{Ja64,SL96}. For relativistic
models, the situation for the instability becomes worse, since
relativistic effects tend to suppress the viscosity-driven instability
(while the CFS-instability becomes stronger). According to recent
results by Bonazzola et al. \cite{BFG97}, for the most relativistic
stars, the viscosity-driven bar mode can become unstable only if
$N<0.55$.  For $1.4 M_{\odot}$ stars, the instability is present for
$N<0.67$.

These results are based on an approximate computation of the
instability in which one perturbs an axisymmetric and stationary
configuration and studies its evolution by constructing a series of
triaxial quasi-equilibrium configurations. During the evolution only
the dominant nonaxisymmetric terms are taken into account.  The method
presented in \cite{BFG97} is an improvement (taking into account
nonaxisymmetric terms of higher order) of an earlier method by the
same authors \cite{BFG96}. Although the method is approximate, its
results indicate that the viscosity-driven instability is likely to be
absent in most relativistic stars, unless the EOS turns out to be
unexpectedly stiff.  

An investigation of the viscosity-driven bar mode instability, using
incompressible, uniformly rotating triaxial ellipsoids in the
post-Newtonian approximation, by Shapiro and Zane \cite{SZ97}, finds
that the relativistic effects increase the critical $T/W$ ratio for
the onset of the instability significantly. More recently, new
post-Newtonian \cite{DiGirolamo02} and fully relativistic calculations
for uniform-density stars \cite{Gondek02} show that the
viscosity-driven instability is not as strongly suppressed by
relativistic effects as suggested in \cite{SZ97}. The most promising
case for the onset of the viscosity-driven instability (in terms of
the critical rotation rate) would be rapidly rotating strange stars
\cite{Gondek03}, but the instability can only appear if its growth
rate is larger than the damping rate due to the emission of
gravitational radiation - a corresponding detailed comparison is still
missing.

\section{Rotating Stars in Numerical Relativity}

Recently, the dynamical evolution of rapidly rotating stars has become
possible in numerical relativity.  In the framework of the 3+1 split
of the Einstein equations \cite{Smarr78} a stationary axisymmetric
star can be  described by a metric of the standard form 
\begin{equation}
ds^2=-(\alpha^2-\beta_i\beta^i) dt^2+2\beta_idx^idt+\gamma_{ij}dx^idx^j,
\end{equation}
where $\alpha$ is the lapse function, $\beta^i$ is the shift three-vector and
$\gamma_{ij}$ is the spatial three-metric, with $i=1\ldots 3$. The spacetime
has the following properties:

\begin{itemize}
\item The metric function $\omega$ in (\ref{e:metric}) describing the
  dragging of inertial frames by rotation is related to the shift
  vector through $\beta^\phi =-\omega $. This shift vector satisfies the
  {\it minimal distortion shift} condition.
  
\item The metric satisfies the {\it maximal slicing} condition, while
  the lapse function is related to the metric function $\nu$ in
  (\ref{e:metric}) through $\alpha =e^\nu$.
  
\item The quasi-isotropic coordinates are suitable for numerical
  evolution, while the radial-gauge coordinates \cite{Bardeen83} are
  not suitable for non-spherical sources (see \cite{BGSM93} for
  details).
  
\item The zero-angular momentum observers (ZAMOs) are the Eulerian
  observers, whose worldlines are normal to the $t$=const.
  hypersurfaces.
  
\item Uniformly rotating stars have $\Omega$=const. in the {\it coordinate
  frame}. This can be shown by requiring a vanishing rate of shear.
  
\item Normal modes of pulsation are discrete in
  the coordinate frame and their frequencies can be obtained by
  Fourier transforms (with respect to coordinate time $t$) of evolved
  variables at a fixed coordinate location \cite{Font00}.
\end{itemize}

Crucial ingredients for the successful long-term evolutions of
rotating stars in numerical relativity are the conformal ADM schemes
for the spacetime evolution (see
\cite{Nakamura87,Shibata95,Baumgarte99,Alcubierre00}) and
hydrodynamical schemes that have been shown to preserve well the sharp
rotational profile at the surface of the star
\cite{Font00,Stergioulas01,Font02}.

\begin{figure}[h]
  \def\epsfsize#1#2{0.6#1} \centerline{\epsfbox{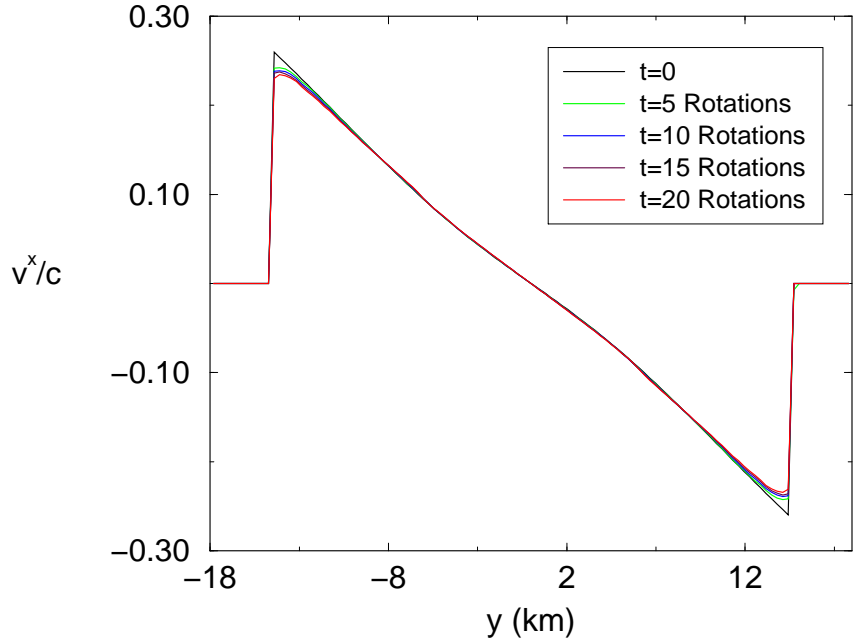}}
  \caption{\it Time-evolution of the rotational velocity profile for a
  stationary, rapidly rotating relativistic star (in the Cowling
  approximation), using the 3rd order PPM scheme and a $116^3$
  grid. The initial rotational profile is preserved to a high degree
  of accuracy, even after 20 rotational periods. (Figure 1 of
Stergioulas and Font, PRL \cite{Stergioulas01}.)}
  \label{fig:rotprof}
\end{figure}

\begin{figure}[p]

  \def\epsfsize#1#2{1#1} \centerline{\epsfbox{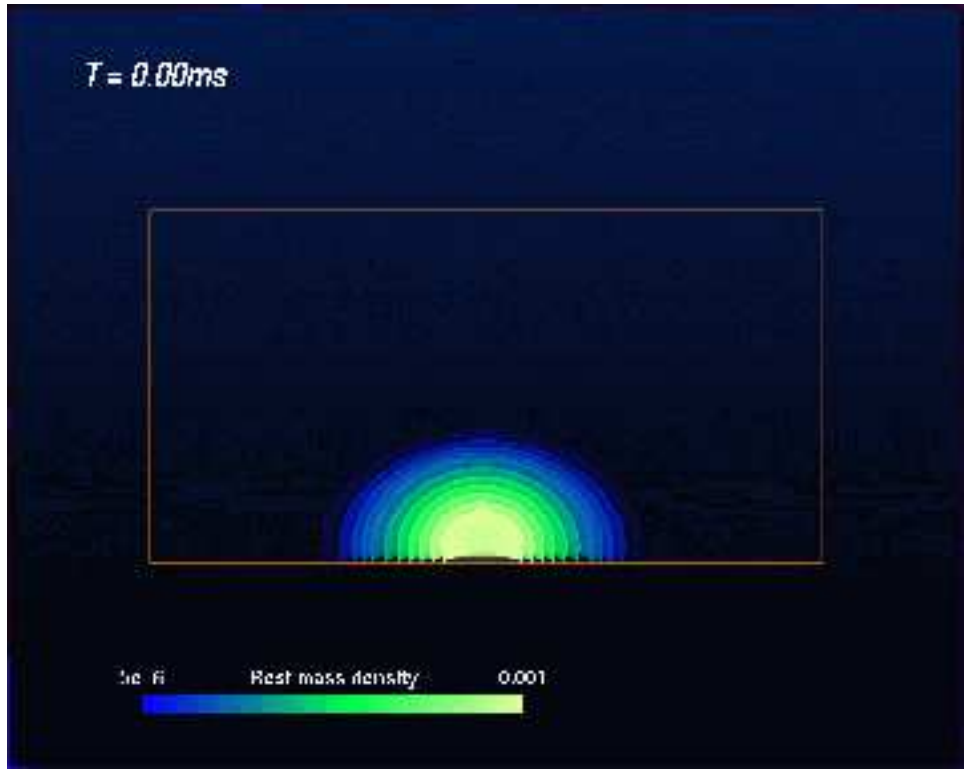}}
  \caption{\it Simulation of a stationary, rapidly rotating neutron
  star model in full General Relativity, for 3 rotational periods
  (shown are iso-density contours, in dimensionless units - the figure
  only shows the initial configuration). The stationary shape is
  well-preserved at a resolution of $129^3$. Simulation by Font,
  Goodale, Iyer, Miller, Rezzolla, Seidel, Stergioulas, Suen and
  Tobias. Visualization by W. Benger and L. Rezzolla at the
  Albert-Einstein-Institute, Golm \cite{AEImovies}.}  \label{fig_rotdens}
\end{figure}
 
\subsection{Numerical Evolution of Equilibrium Models}

\subsubsection{Stable Equilibrium}

The long-term stable evolution of rotating relativistic stars in 3-D
simulations has become possible through the use of
High-Resolution-Shock-Capturing (HRSC) methods (see \cite{Font00b} for
a review). Stergioulas and Font \cite{Stergioulas01} evolve rotating
relativistic stars near the mass-shedding limit for dozens of
rotational periods (evolving only the equations of hydrodynamics) (see Figure
\ref{fig:rotprof}), while accurately preserving the rotational profile,
using the 3rd order PPM method
\cite{Collela84}. This method was shown to be superior to other, commonly
used methods, in 2-D evolutions of rotating relativistic stars
\cite{Font00}.

Fully coupled hydrodynamical and spacetime evolutions in 3-D have been
obtained by Shibata \cite{Shibata99b} and by Font et al.
\cite{Font02}.  In \cite{Shibata99b} the evolution of approximate (conformally
flat) initial data, is presented for about two rotational periods, and
in \cite{Font02} the simulations extend to several full rotational
periods, using numerically exact initial data and a monotonized
central difference (MC) slope limiter
\cite{vanLeer77}.  The MC slope limiter is somewhat less accurate in
preserving the rotational profile of equilibrium stars, than the 3rd
order PPM method, but, on the other hand, it is easier to implement in
a numerical code.

New evolutions of uniformly and differentially rotating stars in 3-D, using
different gauges and coordinate systems, are presented in \cite{Duez03}, while
new 2-D evolutions are presented in \cite{Shibata03}.

\subsubsection{Instability to Collapse}

Shibata, Baumgarte and Shapiro \cite{Shibata00b} study the stability
of supramassive neutron stars rotating at the mass-shedding limit, for
a $\Gamma=2$ polytropic EOS. Their 3-D simulations in full general
relativity show that stars on the mass-shedding sequence with central
energy density somewhat larger than that of the maximum mass model,
are dynamically unstable to collapse.  Thus, the dynamical instability
of rotating neutron stars to axisymmetric perturbations is close to
the corresponding secular instability. The initial data for these
simulations are approximate, conformally flat axisymmetric solutions,
but their properties are not very different from exact axisymmetric
solutions even near the mass-shedding limit \cite{Cook96}. It should
be noted that the approximate minimal distortion (AMD) shift condition
does not prove useful in the numerical evolution, once a horizon
forms.  Instead, modified shift conditions are used in
\cite{Shibata00b}. In the above simulations, no massive disk around
the black hole is formed, as the equatorial radius of the initial
model is inside the radius which becomes the ISCO of the final black
hole. This could change if a different EOS is chosen.

\subsubsection{Dynamical Bar-Mode Instability}
\label{s:dynamical}

\begin{figure}[p]
  \def\epsfsize#1#2{0.6#1} \centerline{\epsfbox{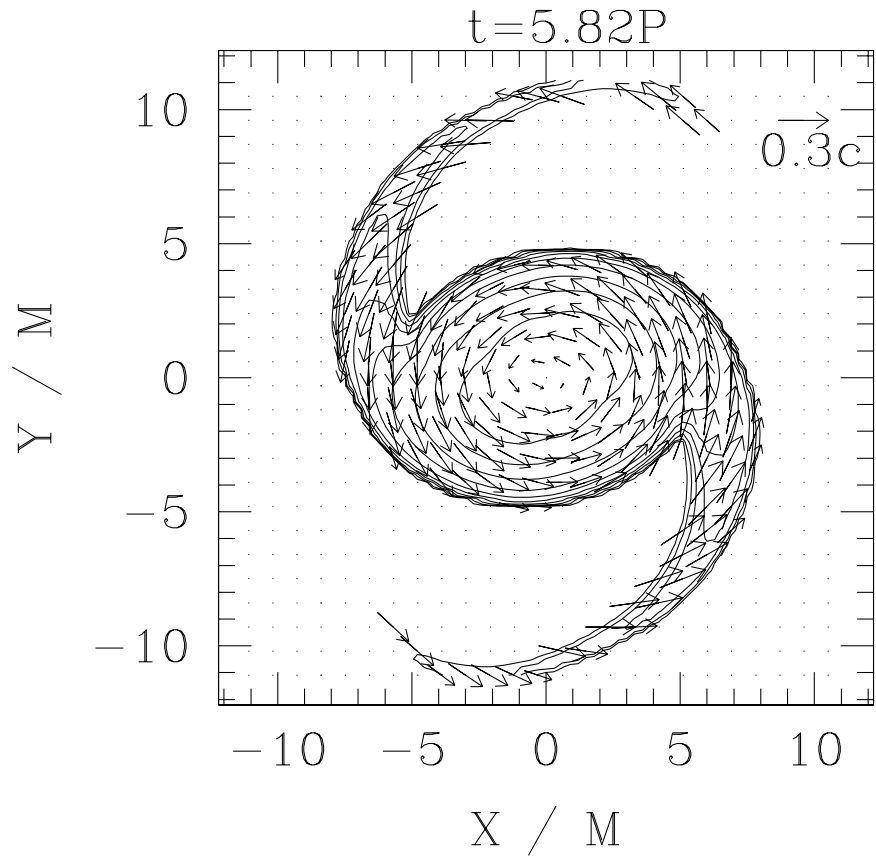}}
  \caption{\it Density contours and velocity flow for a neutron star
  model that has developed spiral arms, due to the dynamical bar-mode
  instability. The computation was done in full General
  Relativity. (Figure 4 of Shibata, Baumgarte, and Shapiro
  in ApJ \cite{Shibata00c}; used with permission).}  \label{fig-spiral}
\end{figure}

\begin{figure}[p]
  \def\epsfsize#1#2{1#1}
  \centerline{\epsfbox{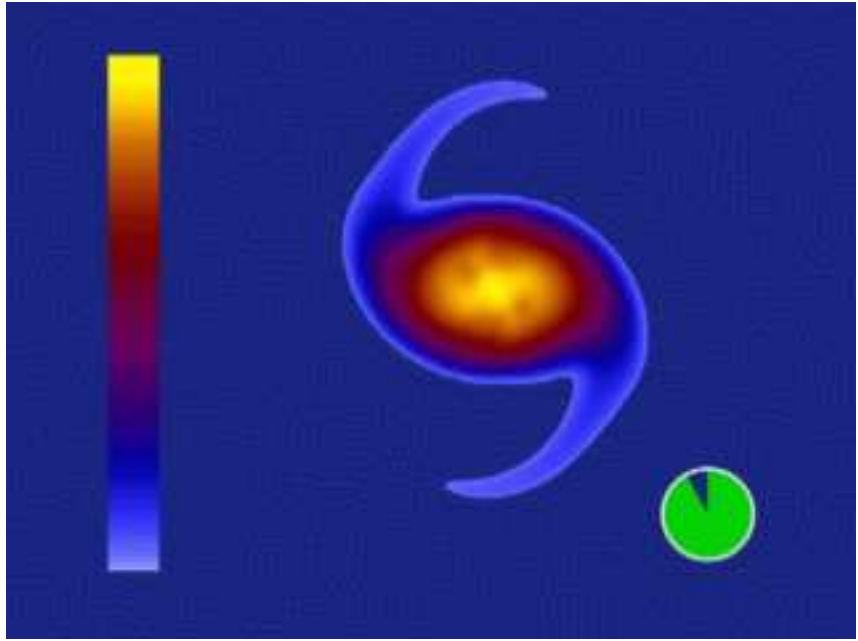}} \caption{\it
  Simulation of the development of the dynamical bar-mode instability
  in a rapidly rotating relativistic star. Spiral arms form within a
  few rotational periods.  The different colors correspond to
  different values of the density, while the computation was done in
  full General Relativity. Movie produced at the University of
  Illinois by Thomas W. Baumgarte, Stuart L. Shapiro and Masaru
  Shibata, with the assistance of the Illinois Undergraduate Research
  Team \cite{Shapiro02}; used with permission.}
  \label{fig-spiraldens}
\end{figure}

\begin{figure}[p]
  \def\epsfsize#1#2{1#1} \centerline{\epsfbox{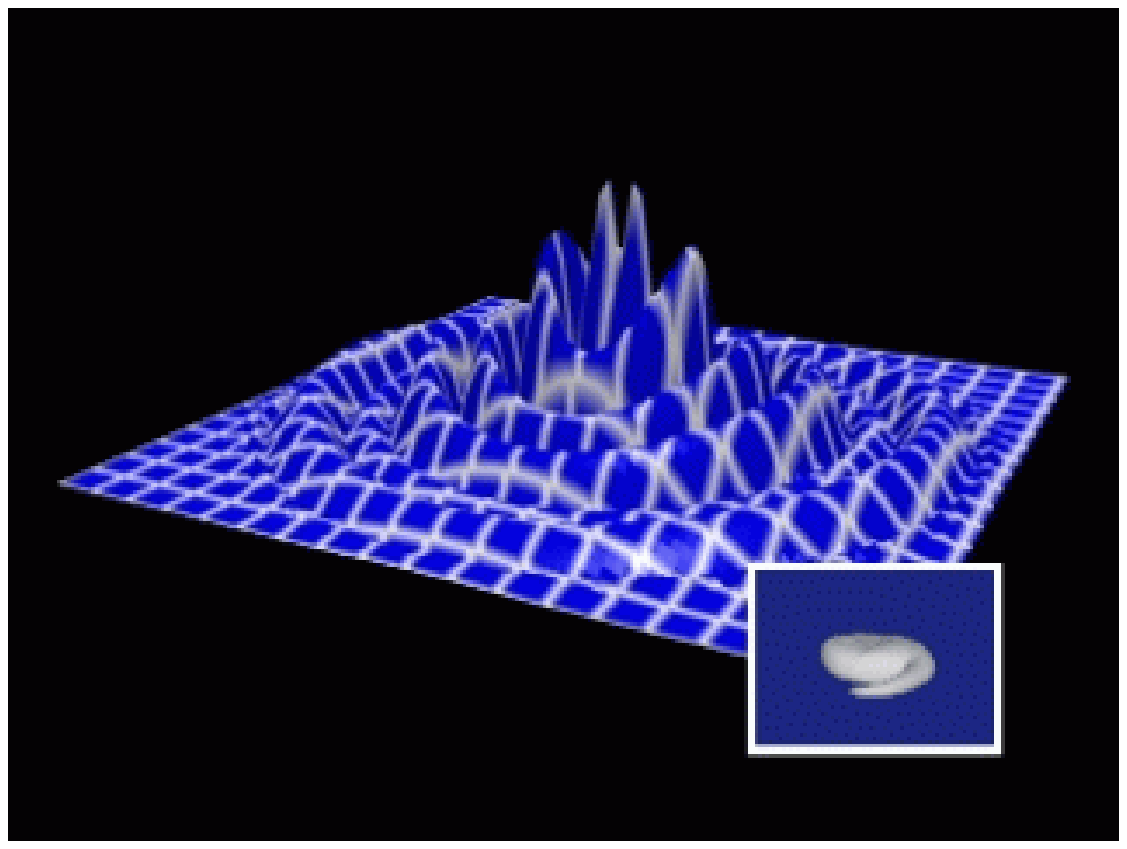}}
  \caption{\it Gravitational wave emission during the development of
  the dynamical bar-mode instability in a rapidly rotating
  relativistic star.  The gravitational wave amplitude in a plane
  containing the rotation axis is shown. At large distances, the waves
  assume a quadrupole-like angular dependence. Movie produced at the
  University of Illinois by Thomas W. Baumgarte, Stuart L. Shapiro and
  Masaru Shibata, with the assistance of the Illinois Undergraduate
  Research Team \cite{Shapiro02}; used with permission.}
  \label{fig-spiralgrav}
\end{figure}

Shibata, Baumgarte and Shapiro \cite{Shibata00c} study the dynamical
bar-mode instability in differentially rotating neutron stars, in
fully relativistic 3-D simulations. They find that stars become
unstable when rotating faster than a critical value of $\beta \equiv T/W \sim
0.24-0.25$. This is only somewhat smaller than the Newtonian value of
$\beta\sim 0.27$. Models with rotation only somewhat above critical become
differentially rotating ellipsoids, while models with $\beta$ much larger
than critical also form spiral arms, leading to mass ejection (see
Figures \ref{fig-spiral}, \ref{fig-spiraldens} and
\ref{fig-spiralgrav}). In any case, the differentially rotating
ellipsoids formed during the bar-mode instability have $\beta >0.2$
indicating that they will be secularly unstable to bar-mode formation,
driven by gravitational radiation or viscosity. The decrease of the
critical value of $\beta$ for dynamical bar formation due to relativistic
effects has been confirmed by post-Newtonian simulations
\cite{Saijo01}.

\subsection{Pulsations of Rotating Stars}
\label{pulsrot}

\begin{figure}[p]
  \def\epsfsize#1#2{0.7#1}
  \centerline{\epsfbox{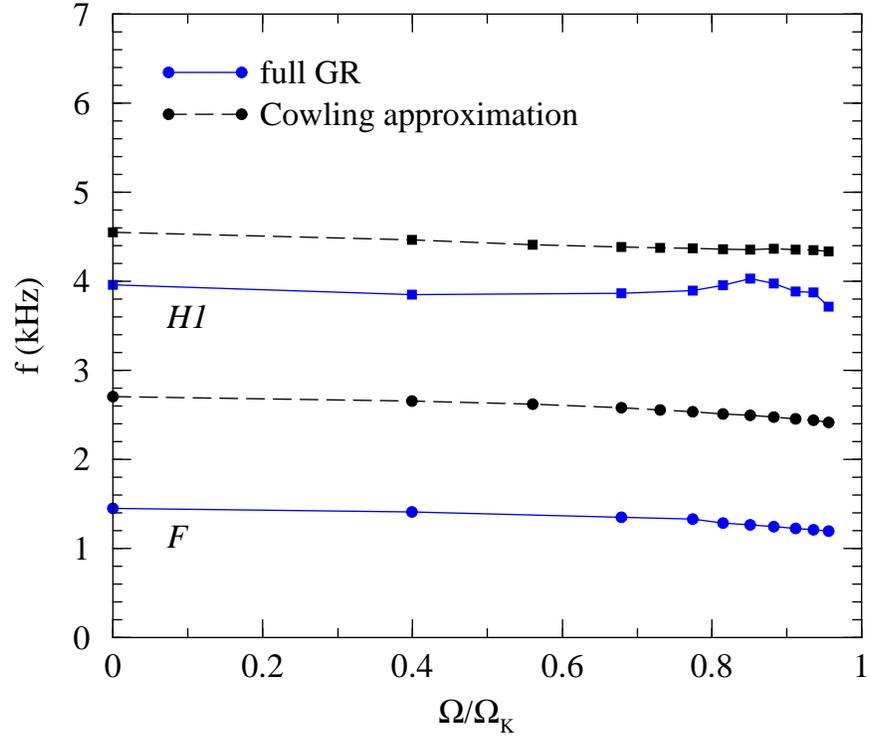}} \caption{\it
  The first fully-relativistic, quasi-radial pulsation frequencies for
  a sequence of rapidly rotating stars (solid lines). The frequencies
  of the fundamental mode $F$ (filled squares) and of the first
  overtone $H_1$ (filled circles) are obtained through {\it coupled}
  hydrodynamical and spacetime evolutions. The corresponding
  frequencies obtained from computations in the relativistic Cowling
  approximation (in \cite{Font01}) are shown as dashed lines. (Figure 16 of
  Font, Goodale, Iyer, Miller, Rezzolla, Seidel, Stergioulas, Suen and 
  Tobias, PRD \cite{Font02}.)} \label{fig:last}
\end{figure}

Pulsations of rotating relativistic stars are traditionally studied
(when possible) as a time-independent, linear eigenvalue problem, but
recent advances in numerical relativity also allow the study of such
pulsations via numerical time-evolutions. The first quasi-radial mode
frequencies of rapidly rotating stars in full general relativity have
been recently obtained in \cite{Font02}, something that has not been
achieved yet with the linear perturbation theory. The fundamental
quasi-radial mode in full general relativity has a similar rotational
dependence as in the relativistic Cowling approximation and an
empirical relation between the full GR computation and the Cowling
approximation can be constructed (Figure \ref{fig:last}). For
higher-order modes, apparent intersections of mode sequences near the
mass-shedding limit do not allow for such empirical relations to be
constructed.

In the relativistic Cowling approximation, 2-D time-evolutions have
yielded frequencies for the $l=0$ to $l=3$ axisymmetric modes of
rapidly rotating relativistic polytropes of $N=1.0$ \cite{Font01}. The
higher-order overtones of these modes show characteristic apparent
crossings near mass-shedding (as was observed for the quasi-radial
modes in \cite{Yoshida00}).

Numerical relativity has also enabled the first study of nonlinear
$r$-modes in rapidly rotating relativistic stars (in the Cowling
approximation) by Stergioulas and Font \cite{Stergioulas01}. For
several dozen dynamical timescales, the study shows that nonlinear
$r$-modes with amplitudes of order unity can exist in a star rotating
near mass-shedding. However, on longer timescales, nonlinear effects
may limit the $r$-mode amplitude to smaller values (see Section
\ref{s_axial}).

\subsection{Rotating Core Collapse}

\subsubsection{Collapse to a Rotating Black Hole}

Black hole formation in relativistic core collapse was first studied
in axisymmetry by Nakamura \cite{Nakamura81,Nakamura83}, using the
(2+1)+1 formalism \cite{Maeda80}. The outcome of the simulation
depends on the rotational parameter
\begin{equation}
q \equiv J/M^2.
\end{equation}
A rotating black hole is formed, only if $q<1$, indicating that
cosmic censorship holds. Stark and Piran \cite{Stark85,Piran86} use
the 3+1 formalism and the radial gauge of Bardeen-Piran
\cite{Bardeen83} to study black hole formation and gravitational wave
emission in axisymmetry. In this gauge, two metric functions used in
determining $g_{\theta\theta}$ and $g_{\phi\phi}$ can be chosen such that they
tend directly at large radii to $h_+$ and $h_\times$ (the even and odd
transverse traceless amplitudes of the gravitational waves, with $1/r$
fall-off at large radii)\footnote{$h_+$ defined in \cite{Stark85} has
  the opposite sign as that commonly used, e.g. in \cite{Thorne83}}.
In this way, the gravitational waveform is obtained at large radii
directly in the numerical evolution. It is also easy to compute the
gravitational energy emitted, as a simple integral over a sphere far
from the source: $\Delta E \sim r^2\int dt(h_{+,r}^2+h_{\times,r}^2)$. Using polar
slicing, black hole formation appears as a region of exponentially
small lapse, when $q<O(1)$. The initial data consists of a
nonrotating, pressure deficient, TOV solution, to which angular
momentum is added by hand. The obtained waveform is nearly independent
of the details of the collapse: it consists of a broad initial peak
(since the star adjusts its initial spherical shape to a flattened
shape, more consistent with the prescribed angular momentum), the main
emission (during the formation of the black hole) and an oscillatory
tail, corresponding to oscillations of the formed black hole
spacetime. The energy of the emitted gravitational waves during the
axisymmetric core collapse is found not to exceed $7\times 10^{-4}M_\odot c^2$ 
(to which the broad initial peak has a negligible contribution). The emitted
energy scales as $q^4$, while the energy in the even mode exceeds by
at least an order of magnitude that in the odd mode.

More recently, Shibata \cite{Shibata00} carried out axisymmetric
simulations of rotating stellar collapse in full general relativity,
using a Cartesian grid, in which axisymmetry is imposed by suitable
boundary conditions.  The details of the formalism (numerical
evolution scheme and gauge) are given in \cite{Shibata99}. It is found
that rapid rotation can prevent prompt black hole formation. When
$q=O(1)$, a prompt collapse to a black hole is prevented even for a
rest mass that is 70-80\% larger than the maximum allowed mass of
spherical stars and this depends weakly on the rotational profile of
the initial configuration. The final configuration is supported
against collapse by the induced differential rotation. In these
axisymmetric simulations, shock formation for $q<0.5$ does not result
in a significant heating of the core: shocks are formed at a
spheroidal shell around the high density core.  In contrast, when the
initial configuration is rapidly rotating ($q= O(1)$), shocks are
formed in a highly nonspherical manner near high density regions and
the resultant shock heating contributes in preventing prompt collapse
to a black hole.  A qualitative analysis in \cite{Shibata00} suggests
that a disk can form around a black hole, during core collapse,
provided the progenitor is nearly rigidly rotating and $q=O(1)$ for a
stiff progenitor EOS. On the other hand, $q<<1$ still allows for a
disk formation, if the progenitor EOS is soft. At present, it is not
clear how much the above conclusions depend on the restriction to
axisymmetry or on other assumptions - 3-dimensional simulations of the
core collapse of such initially axisymmetric configurations have still
to be performed.

A new numerical code for axisymmetric gravitational collapse in the
(2+1)+1 formalism is presented in \cite{Choptuik03}.

\subsubsection{Formation of Rotating Neutron Stars}

First attempts in studying the formation of rotating neutron stars in
axisymmetric collapse were initiated by Evans \cite{Evans84,Evans86}.
Recently, Dimmelmeier, Font \& M{\"u}ller
\cite{Dimmelmeier01,Dimmelmeier01b} have successfully obtained
detailed simulations of neutron star formation in rotating collapse.
In the numerical scheme, HRSC methods are employed for the
hydrodynamical evolution, while for the spacetime evolution the {\it
conformal flatness approximation} \cite{Wilson95} is used.
Surprisingly, the gravitational waves obtained during the neutron star
formation in rotating core collapse are weaker in general relativity
than in Newtonian simulations. The reason for this result is that
relativistic rotating cores bounce at larger central densities than in
the Newtonian limit (for the same initial conditions).  The
gravitational waves are computed from the time-derivatives of the
quadrupole moment, which involves the volume integration of $\rho r^4$.
As the density profile of the formed neutron star is more centrally
condensed than in the Newtonian case, the corresponding gravitational
waves turn out to be weaker. Details of the numerical methods and of
the gravitational wave extraction used in the above studies can be
found in \cite{Dimmelmeier02,Dimmelmeier02b}. 

New, fully relativistic axisymmetric simulations, with coupled
hydrodynamical and spacetime evolution in the light-cone approach,
have been obtained by Siebel et al. \cite{Siebel02,Siebel03} One of
the advantages of the light-cone approach is that gravitational waves
can be extracted accurately at null infinity, without spurious
contamination by boundary conditions. The code by Siebel et al.
combines the light-cone approach for the spacetime evolution with HRSC
methods for the hydrodynamical evolution. In \cite{Siebel03} it is
found that gravitational waves are extracted more accurately using the
Bondi news function, than by a quadrupole formula on the null cone.

A new 2-D code for axisymmetric core collapse, also using HRSC methods
has recently been introduced in \cite{Shibata03}.

\vspace{1cm}
{\bf Acknowledgments}

I am grateful to Emanuele Berti, John L.  Friedman, Wlodek Klu{\'z}niak,
Kostas D.  Kokkotas and Luciano Rezzolla for a careful reading of the
manuscript and for many valuable comments.  Many thanks to Dorota
Gondek-Rosi{\'n}ska and Eric Gourgoulhon for comments and for supplying
numerical results obtained with the Lorene/rotstar code, that were
used in the comparison in Table \ref{Comparison}. I am also grateful
to Marcus Ansorg for discussions and to all authors of the included
figures for granting permission for reproduction. This work was
supported, in part, by the EU Programme ``Improving the Human Research
Potential and the Socio-Economic Knowledge Base'' (Research Training
Network Contract HPRN-CT-2000-00137), KBN-5P03D01721 and the Greek
GSRT Grant EPAN-M.43/2013555.


\bibliography{livrev-02}

\end{document}